\documentclass[letterpaper,journal]{IEEEtran}
\usepackage{amsmath,amsfonts}
\usepackage{algorithmic}
\usepackage{algorithm}
\usepackage{array}
\usepackage{amsthm}
\usepackage[caption=false,font=normalsize,labelfont=sf,textfont=sf]{subfig}
\usepackage{textcomp}
\usepackage{stfloats}
\usepackage{url}
\usepackage{verbatim}
\usepackage{graphicx}
\usepackage{cite}
\usepackage{algorithm}
\usepackage{algorithmic}
\usepackage{amsmath}
\usepackage{amssymb}
\begin{document}

\title{SIDMA: Semantic Interleave Division Multiple Access Communication System}

\author{Yunlu Wang, Chen~Dong$^*$, Sen Wang, Lei Teng, Yaping Sun, Xiaodong Xu,~\IEEEmembership{Senior Member, IEEE},
\\ and Ping Zhang,~\IEEEmembership{Fellow, IEEE}
\thanks{This work is supported in part by the National Key Research and Development Program of China under Grant 2020YFB1806905, in part by the Beijing Natural Science Foundation under Grant L251035, and in part by Beijing University of Posts and Telecommunications-China Mobile
Communications Group Co., Ltd. Joint Institute. \textit{(Corresponding authors: Chen Dong)}

Yunlu Wang, Chen Dong, Lei Teng, Xiaodong Xu and Ping Zhang are with the State Key Laboratory of Networking and Switching Technology, Beijing University of Posts and Telecommunications, Beijing 100876, China. Yunlu Wang is also with the Department of Broadband Communication, Pengcheng Laboratory, Shenzhen 518000, China. (emails: yunlu\_wang@bupt.edu.cn, dongchen@bupt.edu.cn, tenglei@bupt.edu.cn, xuxiaodong@bupt.edu.cn, pzhang@bupt.edu.cn)

Sen Wang is with China Mobile Research Institute, Beijing, China. (email: wangsenyjy@chinamobile.com)

Yaping Sun is with the Department of Broadband Communication, Pengcheng Laboratory, Shenzhen 518000, China. Yaping Sun is also with the Shenzhen Future Network of Intelligence Institute (FNii-Shenzhen), Chinese
University of Hong Kong (Shenzhen), Shenzhen 518172, China. (email: sunyp@pcl.ac.cn)}

}



\maketitle

\begin{abstract}
Multiple Access (MA) technology has consistently served as the core driving force behind the evolution of mobile communications. As a promising paradigm for next-generation communications, Semantic Communication explores entirely new semantic spatial resources by mining the deep meaning of information. However, the inherent spatial correlation and importance heterogeneity of semantic features often cause semantic collisions and semantic collapse in multi-user concurrent transmission scenarios. To address these challenges, this paper proposes a Semantic Interleaved Division Multiple Access (SIDMA) technique. By utilizing a permutation operator to perform structural whitening on semantic features and combining it with an Importance-aware Power Allocation (ImpPA) module for differentiated protection, SIDMA scatters core features across the interleaving domain and adaptively optimizes power levels based on real-time channel conditions. Simulation results demonstrate that, compared with traditional MA techniques and advanced semantic multiple access schemes including Orthogonal-Model Division Multiple Access (OMDMA), Deep Multiple Access (DeepMA), and Shared Embedding (SE), the proposed SIDMA exhibits superior reconstruction fidelity and scalability in multi-user concurrent transmissions, effectively enhancing the communication quality and robustness in resource-constrained environments.
\end{abstract}

\section{Introduction}
\IEEEPARstart{S}{emantic} Communication represents a fundamental shift in the communication paradigm, moving from traditional ``bit transmission'' to ``semantic exchange'' \cite{ref6,gao2025wearable,wei2024toward}. By extracting key semantic information and transmitting only essential features, semantic communication explores entirely new semantic spatial resources by mining the deep meaning of information. This approach demonstrates immense potential to surpass the Shannon limit, particularly in extremely low Signal-to-Noise Ratio (SNR) environments where precise physical bit recovery is impractical. By ensuring the reliable exchange of core meaning, this paradigm provides a pathway toward more intelligent, goal-oriented wireless networks that prioritize information essence over binary accuracy \cite{ref6, sdmcm}.

Throughout this evolution, MA technology has consistently served as the primary driving force, focusing on the efficient and fair sharing of limited physical resources among multiple users \cite{ref1}. From the first to the fourth generation (1G-4G), MA technologies primarily adhered to the guiding principle of orthogonal resource allocation. Starting from Frequency Division Multiple Access (FDMA)\cite{fdma} in the 1G era based on physical band slicing, to Time Division Multiple Access (TDMA)\cite{tdma} in 2G with time-domain segmentation, to Code Division Multiple Access (CDMA)\cite{cdma} in 3G utilizing orthogonal spreading codes for logical isolation, and finally to Orthogonal Frequency Division Multiple Access (OFDMA)\cite{ofdma} in 4G centered on subcarrier orthogonality, the design logic focused on creating ``hard boundaries'' across time, frequency, and code domains to ensure mutual non-interference between user signals \cite{ref2}. Entering the 5G era, research shifted toward non-orthogonal resource reuse, notably with Non-Orthogonal Multiple Access (NOMA)\cite{noma} using power-domain multiplexing and Successive Interference Cancellation (SIC) \cite{ref3}. Simultaneously, Rate-Splitting Multiple Access (RSMA)\cite{rsma} emerged as a more robust and universal framework \cite{ref4, ref5}, enabling flexible switching between orthogonal and non-orthogonal modes to enhance connectivity and reliability.

As communication research deepens, the integration of semantic extraction with multiple access—termed Semantic Multiple Access—has emerged as a vital research direction \cite{omdma}. Recent explorations have unfolded across three primary dimensions. First, AI-native multiple access paradigms have been proposed to exploit model-level or feature-level resources, such as Orthogonal-Model Division Multiple Access (OMDMA) \cite{omdma} and Deep Multiple Access (DeepMA)\cite{zhang2024deepma}. These schemes attempt to separate users in high-dimensional semantic spaces through orthogonal model subspaces or mutually orthogonal symbol vectors. Second, research has focused on integrating traditional NOMA and its variants with semantic extraction to optimize resource allocation. For instance, NOMA-enhanced architectures have been developed to capture long-range dependencies and optimize superposition coding based on semantic differences \cite{li2025noma}. Third, RSMA exhibits great potential in managing interference within heterogeneous semantic networks due to its dynamic precoding and message-splitting flexibility \cite{liu2026rate}. These advancements aim to achieve deep coupling between semantic priorities and physical resources, providing a unified framework for multi-user semantic exchange.

However, while traditional MA technologies have achieved significant efficiency gains at the bit level\cite{wang2025semanticsurvey}, their design logic remains confined to the precise recovery of every physical bit. Consequently, these methods fail to address the underlying semantic correlations within the information. Furthermore, most existing semantic multiple access schemes continue to perform resource scheduling based primarily on physical-layer attributes such as signal power or channel matrices. As a result, they overlook unique semantic dimensions including the heterogeneity of importance distribution and non-uniform interference characteristics. Due to the inherent high spatial correlation of semantic features, non-orthogonal superposition in these schemes often leads to serious semantic collisions. For instance, key semantic blocks from different users, such as high-value features extracted by a Swin Transformer, may overlap at the same physical resource location. This overlap leads to a catastrophic loss of detail, a phenomenon known as semantic collapse. Such degradation remains inevitable even with the application of complex interference cancellation algorithms \cite{lee2026transformer}.

To address these challenges, inspired by the IDMA \cite{idma} scheme which utilizes independently and randomly generated interleavers to disperse coded sequences and minimize correlation, this paper proposes a Semantic Interleaving Division Multiple Access (SIDMA) architecture. By introducing a permutation operator to perform ``structural whitening'' on semantic features—effectively extending the concept of chip-level interleaving to the semantic feature domain—and combining it with an adaptive power allocation module, this framework achieves differentiated protection for core information. The main contributions of this paper are summarized as follows:

\begin{itemize}
\item \textbf{Design of the SIDMA System Architecture:} A comprehensive end-to-end SIDMA architecture is developed and implemented, covering modules from semantic encoding and symbol-level interleaving to de-interleaving and decoding. The core innovation lies in utilizing a permutation operator to perform structural whitening on the original semantic feature maps. This process effectively breaks the spatial correlation of features, establishing a unified framework for efficient multi-user semantic multiplexing and interference mitigation. 
\item \textbf{Revealing Semantic Whitening and Asymptotic Orthogonality Mechanisms:} The effectiveness of the semantic interleaving module is theoretically demonstrated. By introducing the Peak Sidelobe Level (PSL) metric, the study proves that random permutation allows the cross-correlation properties of semantic vectors to converge to zero in probability, exhibiting asymptotic orthogonality in high-dimensional feature spaces. 
\item \textbf{Importance-Aware Adaptive Power Allocation Strategy:} An importance-aware adaptive power allocation (ImpPA) module is proposed to jointly perceive the semantic importance of image content and real-time channel conditions. This strategy provides differentiated protection for key semantic information, effectively avoiding collision losses of high-value features and significantly enhancing reconstruction gains.
\item \textbf{Comprehensive Simulation and Ablation Analysis:} Extensive simulations demonstrate that SIDMA significantly outperforms traditional methods, OMDMA \cite{omdma}, DeepMA \cite{zhang2024deepma}, and SE \cite{lee2026transformer} in terms of peak signal-to-noise ratio (PSNR) and structural similarity (SSIM). Notably, the proposed system successfully scales to support up to 100 concurrent users—which, to the best of our knowledge, is the first time such massive connectivity has been achieved in the field of semantic multiple access. Furthermore, systematic ablation experiments quantify the superiority of the ImpPA module, validating the advanced nature of the proposed framework.
\end{itemize}

\section{Related Works}
Existing studies on multi-user semantic communications have primarily focused on integrating traditional NOMA and its variants with semantic extraction. For instance, in Satellite-Terrestrial Integrated Networks, a NOMA-enhanced semantic architecture leveraged the Swin Transformer’s ability to capture long-range dependencies \cite{li2025noma}. To further mitigate semantic distortion, the KDD-SemNOMA framework combined knowledge distillation with diffusion models to ensure robust feature transmission \cite{wang2026knowledge}. Furthermore, a semantics-empowered NOMA framework, termed CIS-MA, was proposed to exploit the correlation between multiple users by extracting common information based on Wyner's information theory, which utilizes a two-stage signal detection and decoding strategy to enhance reconstruction quality\cite{li2025semanticsnoma}. Beyond the power domain, Sparse Code Multiple Access (SCMA) has been introduced to semantic systems, where its hierarchical transmission is effectively integrated with semantic importance levels and optimized codebook designs to enhance capacity \cite{wang2025semantic}.

Furthermore, Rate-Splitting Multiple Access (RSMA) exhibits great potential in heterogeneous semantic networks due to its interference management flexibility. The APVST framework integrates RSMA with deep JSCC to handle overlapping Fields of View (FoV) in panoramic video transmission \cite{gao2025rate}. For probabilistic semantic communication (PSCom), researchers utilized knowledge graphs and rate-splitting to bridge computation and communication, balancing energy consumption through optimized semantic compression ratios (SCR) \cite{zhao2025compression, xu2025rate}. Moreover, the coexistence of semantic and bit communications under RSMA has been investigated, revealing its superiority over NOMA in high semantic rate regimes \cite{liu2026rate}. Despite these advances in power and code domains, the dynamic reshaping of interference via the interleaving domain remains unexplored.

Recently, AI-native multiple access paradigms have emerged to exploit semantic-level resources. OMDMA were proposed to serve users with distinct or orthogonal semantic models \cite{omdma}. The DeepMA framework achieved end-to-end multiple access by encoding data into mutually orthogonal semantic symbol vectors \cite{zhang2024deepma}. In the feature domain, Semantic Feature Division Multiple Access (SFDMA) utilizes separate feature subspaces to minimize interference \cite{ma2025semantic, shen2026semantic}. Token-Domain Multiple Access (ToDMA) employs Transformer-generated tokens as universal semantic units to mitigate collisions \cite{qiao2025token}, while the Shared Embedding (SE) scheme separates users in high-dimensional spaces using learnable positional masks and attention-driven demultiplexing \cite{lee2026transformer}. Additionally, Semantic Feature Multiple Access (SFMA) optimizes resource allocation in GAI-empowered networks by introducing semantic weight parameters \cite{wang2025generative, wang2025sfma_infocom}. While these paradigms advance AI-native multiplexing, they overlook the exploitation of the interleaving domain for structural whitening, which motivates our proposed SIDMA.

\section{SIDMA Based Semantic Communication System}
This section details the proposed Semantic Interleaving Division Multiple Access (SIDMA) system. SIDMA addresses semantic collisions in multi-user scenarios by coupling feature-domain ``structural reshaping'' with physical-layer ``adaptive scheduling.'' The system model is established in Section III-A, followed by an analysis of the ``structural whitening'' mechanism in Section III-B. An importance-aware adaptive power allocation strategy is proposed in Section III-C. Section III-D introduces an optimal seed selection algorithm based on the Peak Sidelobe Level (PSL), and the specific network implementation is described in Section III-E.
\subsection{System Model}

\begin{figure*}[t]
    \centering
    \includegraphics[width=0.95\textwidth]{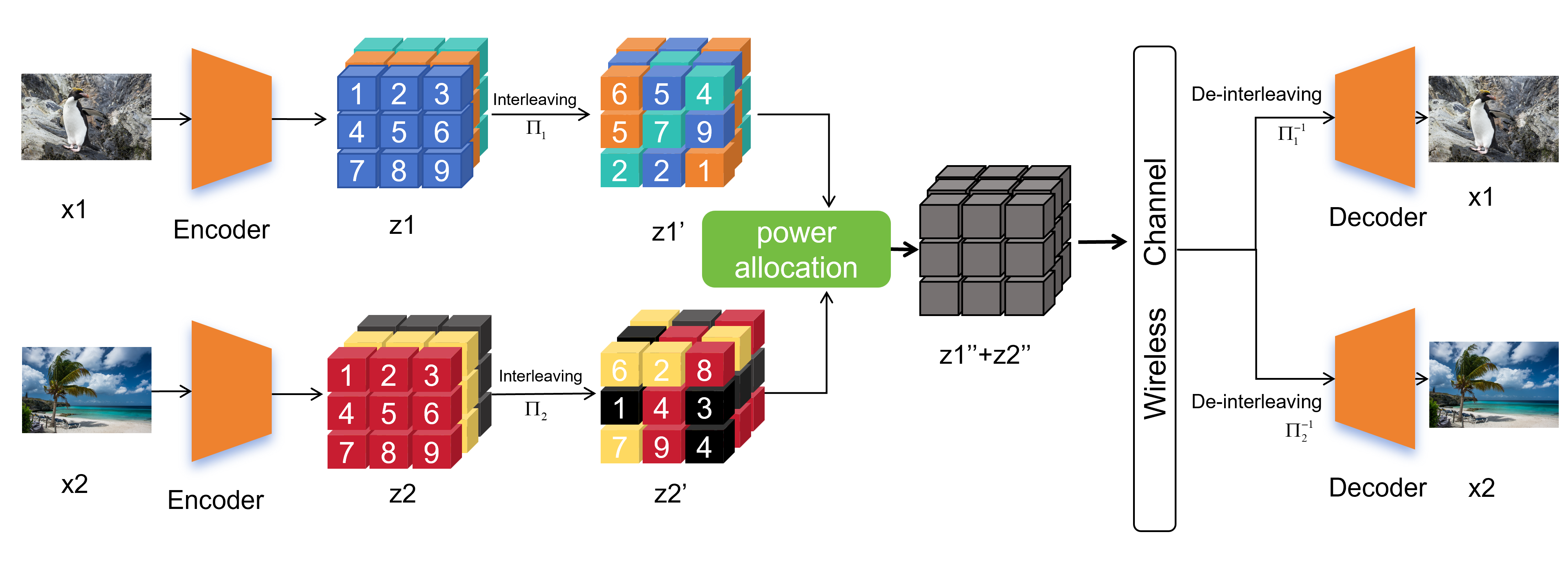}
    \caption{The overall framework of the proposed Semantic Interleaving Division Multiple Access (SIDMA) system for multi-user downlink transmission, comprising semantic encoding, importance-aware power allocation (ImpPA), semantic interleaving, and joint decoding modules.}
    \label{fig:system_framework}
\end{figure*}

This subsection presents the framework of the proposed SIDMA system. As illustrated in Fig. \ref{fig:system_framework}, a multi-user wireless communication scenario is considered where $K$ users share the same time-frequency resources. For each user $k \in \{1, \dots, K\}$, the source image is denoted by $\mathbf{s}_k \in \mathbb{R}^{H \times W \times 3}$.
The source image $\mathbf{s}_k$ is first processed by a semantic encoder $E_{\theta}$, implemented via a Swin Transformer-based architecture, to extract latent semantic features:
\begin{equation}
\mathbf{z}_k = E_{\theta}(\mathbf{s}_k),
\end{equation}
where the dimensions of the extracted feature map are given by $\mathbf{z}_k \in \mathbb{R}^{h \times w \times C}$.
Due to the hierarchical attention mechanism, these features exhibit significant spatial importance heterogeneity. To provide differentiated protection for core semantic information, an importance-aware adaptive power allocation module is utilized. This module perceives the image content and generates a power weighting matrix $\mathbf{P}_k \in \mathbb{R}^{h \times w \times C}$ by mapping semantic saliency and channel conditions to specific power levels. The weighted semantic features $\tilde{\mathbf{z}}_k$ are then obtained through the following operation:
\begin{equation}
\tilde{\mathbf{z}}_k = \mathbf{P}_k \odot \mathbf{z}_k,
\end{equation}
where $\odot$ denotes the element-wise product. The power matrix $\mathbf{P}_k$ is subject to the total power constraint:
\begin{equation}
\frac{1}{hwC} \sum_{i=1}^{h} \sum_{j=1}^{w} \sum_{c=1}^{C} \left|P_{k,(i,j,c)}\right|^2 \leq P_{\text{total}}.
\end{equation}
To mitigate semantic collisions during non-orthogonal multiplexing, each user applies a unique permutation operator $\Pi_k$ to the weighted features. The resulting interleaved semantic symbols $\mathbf{x}_k$ are expressed as:
\begin{equation}
\mathbf{x}_k = \Pi_k(\tilde{\mathbf{z}}_k).
\end{equation}
Expanding the entire encoding and modulation process, the transmitted signal is represented as:
\begin{equation}
\mathbf{x}_k = \Pi_k \left( \mathbf{P}_k \odot E_{\theta}(\mathbf{s}_k) \right).
\end{equation}
The interleaved signals from all $K$ users are then superposed and transmitted over the wireless channel. The received signal $\mathbf{y}$ is modeled as:
\begin{equation}
\mathbf{y} = \sum_{k=1}^{K} h_k \mathbf{x}_k + \mathbf{n},
\end{equation}
where $h_k$ represents the complex channel gain for user $k$, and the additive white Gaussian noise (AWGN) $\mathbf{n}$ follows the distribution $\mathcal{CN}(0, \sigma^2 \mathbf{I})$. At the receiver side, the objective is to recover the source image for each user. For a target user $k$, the receiver applies the corresponding inverse permutation operator $\Pi_k^{-1}$, yielding the de-interleaved features:
\begin{equation}
\hat{\mathbf{z}}_k = \Pi_k^{-1}(\mathbf{y}).
\end{equation}
Substituting the channel model, the de-interleaved features are decomposed into the target signal and the reshaped interference:
\begin{equation}
\hat{\mathbf{z}}_k = h_k \tilde{\mathbf{z}}_k + \Pi_k^{-1} \left( \sum_{j \neq k} h_j \mathbf{x}_j + \mathbf{n} \right).
\end{equation}
Finally, these features are fed into a semantic decoder $D_{\phi}$ to reconstruct the image:
\begin{equation}
\hat{\mathbf{s}}_k = D_{\phi}(\hat{\mathbf{z}}_k).
\end{equation}

\subsection{Mechanism of Semantic Structural Whitening}
The core philosophy of SIDMA is to transform destructive ``semantic collisions'' into manageable ``unstructured noise.'' In conventional non-orthogonal systems, semantic features extracted by hierarchical architectures exhibit strong structural correlations, which leads to catastrophic interference when multiple users share the same physical resources. Such structured interference often misguides the neural decoder into synthesizing hallucinatory artifacts, ultimately resulting in semantic collapse.

To distinguish the proposed approach from the basic system flow, the interaction between different user operators is analyzed. In SIDMA, the interleaving process redefines the multi-user interaction. After the de-interleaving process at the receiver, the target user's signal is restored. However, the interference from any other user $j$ is processed by a mismatched operator pair. The \textit{composite scrambling operator} $\Phi_{j \to k}$ is defined as:
\begin{equation}\label{eq:composite_op}
    \Phi_{j \to k} \triangleq \Pi_k^{-1} \circ \Pi_j, \quad \text{for } j \neq k.
\end{equation}
This operator represents the ``effective channel'' seen by the interference signal in the feature domain.

Since $\Pi_k$ and $\Pi_j$ are independently chosen from a high-dimensional permutation space, the composite operator $\Phi_{j \to k}$ functions as a random spatial decorrelator. It maps the spatially correlated interference $\tilde{\mathbf{z}}_j$ to a scrambled sequence:
\begin{equation}
    \mathbf{I}_{j \to k} = \Phi_{j \to k}(\tilde{\mathbf{z}}_j).
\end{equation}
The effect of $\Phi_{j \to k}$ is to ``shatter'' the original semantic topology of the interference. Even if the original features contained structured objects, $\mathbf{I}_{j \to k}$ manifests as a collection of non-coherent, pseudo-random elements.

This mechanism exploits a key property of hierarchical decoders: they are designed to prioritize structured semantic patterns while treating unstructured high-frequency components as noise. By ensuring that the composite operator $\Phi_{j \to k}$ produces a near-white interference distribution, SIDMA allows the decoder to perform effective denoising:
\begin{equation}
    \hat{\mathbf{s}}_k = D_{\phi} \left( h_k \tilde{\mathbf{z}}_k + \text{Noise-like}(\mathbf{I}_{j \to k}) \right).
\end{equation}
Thus, the system achieves multiplexing gain not by avoiding collisions, but by ensuring that collisions are ``semantically uninformative'' to the decoder. The statistical convergence of this scrambling effect is rigorously analyzed in Section IV.

\subsection{Importance-aware Adaptive Power Allocation}
In the SIDMA system, simply interleaving semantic features is insufficient to guarantee optimal reconstruction, especially when total transmission power is limited. Since different elements in the semantic feature map $\mathbf{z}_k$ contribute unequally to global image understanding, an Importance-aware Power Allocation (ImpPA) strategy is proposed to provide differentiated protection for high-value information.

The importance of each semantic element is first quantified through a saliency extraction process. Given latent features $\mathbf{z}_k$, an importance map $\mathbf{M}_k$ is derived from the attention weights of the Swin Transformer:
\begin{equation}
\mathbf{M}_k = \mathcal{A}(\mathbf{z}_k),
\end{equation}
where $\mathcal{A}(\cdot)$ denotes the importance evaluation function. Elements in $\mathbf{M}_k \in \mathbb{R}^{h \times w \times C}$ reflect the sensitivity of reconstruction quality to distortion of corresponding feature components.

To dynamically map importance values to physical power gains, an adaptive power allocation module is employed. This module perceives the current importance map and signal-to-noise ratio (SNR) to generate a power weighting matrix:
\begin{equation}
\mathbf{P}_k = f_{\mathrm{alloc}}(\mathbf{M}_k,\gamma_k),
\end{equation}
where $f_{\mathrm{alloc}}(\cdot)$ represents the adaptive mapping function, and $\gamma_k$ is the instantaneous signal-to-noise ratio (SNR) of user $k$. The adaptive gain for each semantic element $(i,j,c)$ is defined as:
\begin{equation}
P_{k,(i,j,c)} = \sqrt{\frac{P_{\text{total}}\,\beta_{k,(i,j,c)}}{\frac{1}{hwC}\sum_{i,j,c}\beta_{k,(i,j,c)}}}.
\end{equation}
where $\beta_{k,(i,j,c)}$ is the importance-dependent scaling factor output by the adaptive power allocation module. This allocation strategy ensures that
\begin{equation}
P_{k,(i,j,c)} > P_{k,(l,m,n)} \quad \text{if} \quad M_{k,(i,j,c)} > M_{k,(l,m,n)}.
\end{equation}

The objective of ImpPA is to maximize the effective SINR for high-importance semantic blocks during non-orthogonal multiplexing. In the presence of multi-user interference, the effective SINR for a specific semantic element $(i,j,c)$ is expressed as:
\begin{equation}
\mathrm{SINR}^{\mathrm{eff}}_{k,(i,j,c)} = \frac{|h_k|^2\,|P_{k,(i,j,c)}|^2}{\sum_{j\neq k}|h_j|^2\left|P_{j,\Pi_j\Pi_k^{-1}(i,j,c)}\right|^2 + \sigma^2}.
\end{equation}
By assigning higher power to essential features, the ImpPA mechanism ensures that even if a structural collision occurs in the interleaving domain, high-importance elements can maintain sufficient power advantage over interfering signals. Consequently, the residual distortion after decoding is minimized:
\begin{equation}
\min_{\mathbf{P}_k} \sum_{k=1}^{K} \mathbb{E}\left[\left|\mathbf{s}_k - \hat{\mathbf{s}}_k\right|^2\right].
\end{equation}
Through this content-aware power scheduling, SIDMA avoids semantic collapse of critical information (e.g., edges and textures) while tolerating moderate interference on secondary background features. This differentiated protection mechanism significantly enhances reconstruction fidelity and robustness in resource-constrained multi-user environments.

\subsection{Peak Sidelobe Level (PSL) based Seed Selection}
While random permutations can achieve asymptotic whitening in high-dimensional spaces, finite-dimensional semantic feature maps in practical systems may still exhibit residual structural correlations. To ensure effective interference suppression, a seed-selection mechanism is proposed to minimize the Peak Sidelobe Level (PSL), which quantifies the maximum off-diagonal structural correlation\cite{psl}.

Let $\mathbf{z} \in \mathbb{R}^{N}$ be a vectorized semantic feature map. To establish a consistent baseline for correlation analysis, $\mathbf{z}$ is normalized to possess zero mean and unit variance, such that $\sum_{i=1}^{N} z_i = 0$ and $\frac{1}{N}\sum_{i=1}^{N} z_i^2 = 1$.

For a spatial shift $\Delta \neq 0$, the discrete autocorrelation function is defined as:
\begin{equation}
R(\Delta) = \sum_{i=1}^{N} z_i z_{i+\Delta},
\end{equation}
where index $i+\Delta$ is computed modulo $N$.

The PSL is then defined as the ratio of the maximum absolute sidelobe to the main lobe:
\begin{equation}
PSL = \frac{\max_{\Delta \neq 0} |R(\Delta)|}{R(0)}.
\end{equation}

A lower $PSL$ indicates that permutation $\Pi_s$ more effectively scatters spatial dependencies of semantic features, transforming structured interference into unstructured noise. The selection procedure is summarized in Algorithm~\ref{alg:seed_selection}.

\begin{algorithm}[h]
\caption{PSL-based Interleaving Seed Selection}
\label{alg:seed_selection}
\begin{algorithmic}[1]
\REQUIRE Pre-trained semantic encoder $E_{\theta}$, target image set $\mathcal{D}$, candidate seed set $\mathcal{S}_{\text{cand}}=\{s_1,s_2,\dots,s_N\}$, number of selected seeds $N_{\text{sel}}$.
\ENSURE Top $N_{\text{sel}}$ seeds with minimal structural interference.
\STATE \textbf{Initialization:} Extract feature maps $\mathbf{Z}=\{E_{\theta}(\mathbf{s})\mid \mathbf{s}\in\mathcal{D}\}$.
\STATE Flatten and normalize features to zero mean and unit variance: $\mathbf{v}=\mathrm{Normalize}(\mathrm{vec}(\mathbf{Z}))$.
\FOR{each seed $s \in \mathcal{S}_{\text{cand}}$}
\STATE Generate permutation sequence $\Pi_s$ from seed $s$.
\STATE Apply permutation to normalized features: $\mathbf{x}_s=\Pi_s(\mathbf{v})$.
\STATE Compute discrete autocorrelation $R_s(\Delta)$ for all $\Delta \in \{1,\dots,N-1\}$.
\STATE Calculate PSL score: $PSL(s)=\max_{\Delta\neq 0}|R_s(\Delta)|/R_s(0)$.
\ENDFOR
\STATE Sort $\mathcal{S}_{\text{cand}}$ in ascending order based on $PSL(s)$.
\STATE \textbf{Output:} Select the top $N_{\text{sel}}$ seeds as the optimal set $\mathcal{S}_{\text{opt}}$.
\end{algorithmic}
\end{algorithm}

\subsection{Network Architecture}
\begin{figure*}[!t]
\centering
\includegraphics[width=0.80\textwidth]{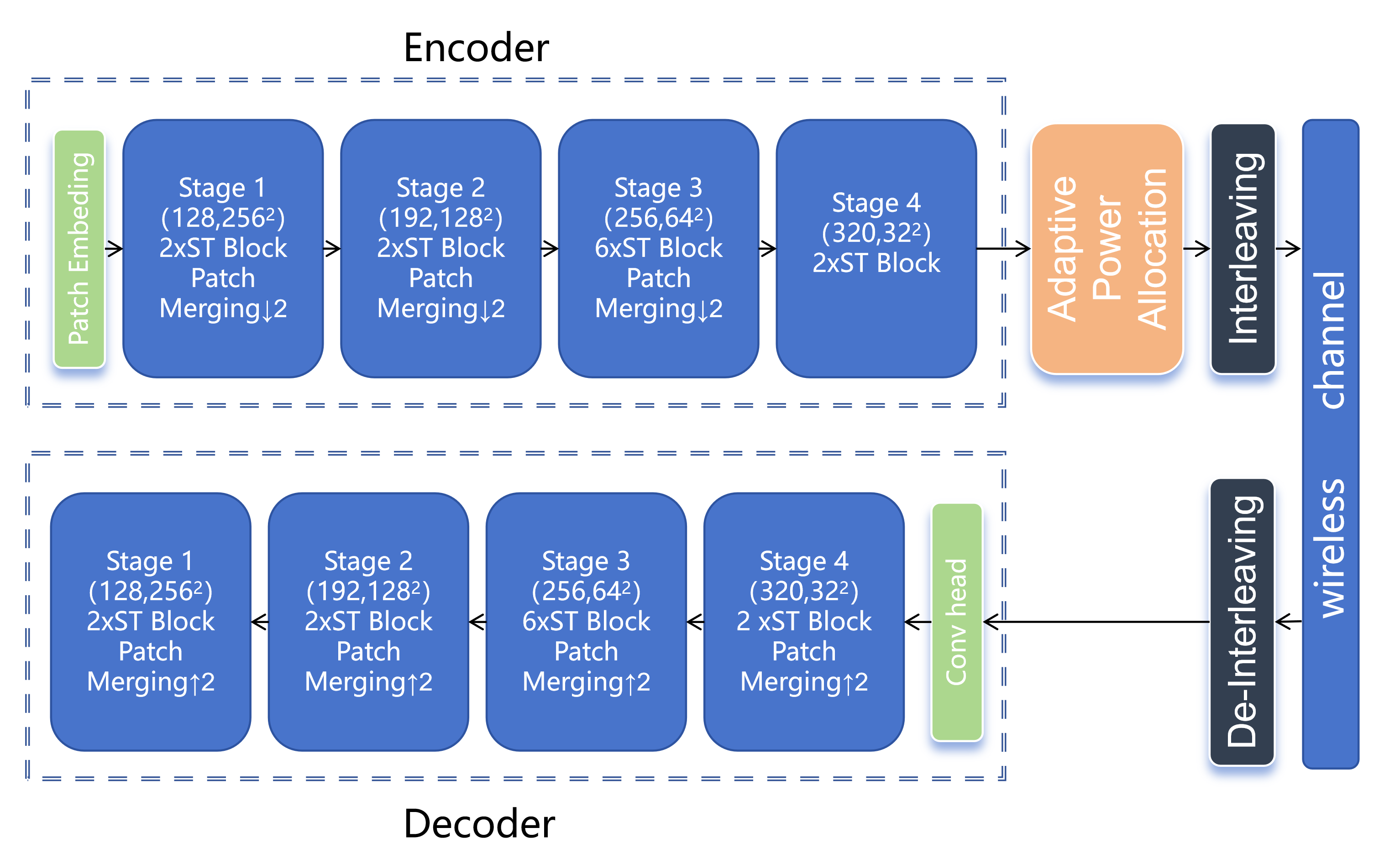}
\caption{The detailed network architecture of SIDMA, featuring a semantic Encoder, a neural network-based adaptive power allocation module, and a semantic Decoder.}
\label{fig:network_architecture}
\end{figure*}
In this subsection, the network architecture of the SIDMA system is presented. The architecture is primarily composed of three functional modules: a Semantic Encoder, a Neural Network-based Adaptive Power Allocation Module, and a Semantic Decoder.

\subsubsection{Semantic Encoder}
The semantic encoder $E_{\theta}$ adopts a hierarchical Swin Transformer backbone to capture multi-scale semantic features. The input image $\mathbf{s}_k$ is first partitioned into non-overlapping patches, and each patch is projected into a $C$-dimensional embedding:
\begin{equation}
\mathbf{X}_0 = \mathrm{Linear}(\mathrm{PatchPartition}(\mathbf{s}_k)).
\end{equation}
The backbone contains multiple stages with Swin Transformer blocks. In the $l$-th block, feature updates are
\begin{equation}
\hat{\mathbf{z}}^{l} = \mathrm{W\text{-}MSA}(\mathrm{LN}(\mathbf{z}^{l-1})) + \mathbf{z}^{l-1},
\end{equation}
\begin{equation}
\mathbf{z}^{l} = \mathrm{MLP}(\mathrm{LN}(\hat{\mathbf{z}}^{l})) + \hat{\mathbf{z}}^{l},
\end{equation}
where $\mathrm{W\text{-}MSA}(\cdot)$ denotes the Window-based Multi-head Self-Attention, which restricts attention computation within local non-overlapping windows to significantly reduce computational complexity. $\mathrm{LN}(\cdot)$ and $\mathrm{MLP}(\cdot)$ represent layer normalization and multilayer perceptron, respectively. The importance map is aggregated from attention scores:
\begin{equation}
\mathbf{M}_k = \sum_{h=1}^{H} \mathrm{Softmax}\left(\frac{\mathbf{Q}_h\mathbf{K}_h^T}{\sqrt{d}} + \mathbf{B}\right),
\end{equation}
where $\mathbf{Q}_h$ and $\mathbf{K}_h$ are the query/key matrices for head $h$, and $\mathbf{B}$ is the relative position bias.

\subsubsection{Neural Network-based Adaptive Power Allocation Module}
The Importance-aware Power Allocation (ImpPA) module is implemented as a dedicated neural network that takes the importance map $\mathbf{M}_k$ and the instantaneous SNR $\gamma_k$ as inputs. To enable joint perception of content and channel, the scalar $\gamma_k$ is broadcast to match the dimensions of the importance map, and the input tensor $\mathbf{U}_k$ is formed as
\begin{equation}
\mathbf{U}_k = [\mathbf{M}_k,\mathbf{\Gamma}_k],
\end{equation}
where $\mathbf{\Gamma}_k$ is the tiled SNR tensor and $[\cdot]$ denotes concatenation along the channel dimension.

Unlike conventional DRL-based agents, this module is a feed-forward neural network $f_{\phi}$ designed to learn the optimal non-linear mapping between semantic significance and power weights. The network output, denoted as raw scaling coefficients $\boldsymbol{\beta}_k$, is given by
\begin{equation}
\boldsymbol{\beta}_k = f_{\phi}(\mathbf{U}_k).
\end{equation}

To satisfy the total power constraint $P_{\text{total}}$, the final adaptive power weighting matrix $\mathbf{P}_k$ is obtained by normalizing $\boldsymbol{\beta}_k$ as
\begin{equation}
P_{k,(i,j,c)} = \sqrt{\frac{P_{\text{total}}\,\beta_{k,(i,j,c)}}{\frac{1}{hwC}\sum_{i,j,c}\beta_{k,(i,j,c)}}}.
\end{equation}
By jointly perceiving content-level importance and channel quality, the network performs fine-grained feature-level power scaling. This ensures that core semantic blocks, such as object contours and textures, are robustly protected against multi-user interference through learned importance-to-power strategies.
\subsubsection{Semantic Decoder}
The decoder $D_{\phi}$ mirrors the encoder and performs inverse operations. After de-interleaving, features are processed through Patch Unmerging and Swin blocks to restore spatial resolution:
\begin{equation}
\hat{\mathbf{z}}^{m} = \mathrm{SwinBlock}(\hat{\mathbf{z}}^{m-1}),
\end{equation}
\begin{equation}
\hat{\mathbf{s}}_k = \mathrm{Conv}(\mathrm{PatchUnmerging}(\hat{\mathbf{z}}^{M})).
\end{equation}
This symmetric design allows high-dimensional semantic correlations captured in encoding to suppress whitened multi-user interference during reconstruction.

\section{Theoretical Analysis}
This section theoretically validates the SIDMA framework by sequentially quantifying the structural whitening of individual semantic features and the asymptotic orthogonality among multiple users.

\subsection{Asymptotic Convergence of Peak Sidelobe Level in SIDMA}
\label{sec:psl_convergence}

Let $\mathcal{Z} = \{z_1, z_2, \dots, z_N\}$ denote the structurally whitened semantic feature set of a given user, where $N$ is the interleaver length. Based on the empirical normalization, the exact zero-mean and unit-variance constraints are:
\begin{equation}
\begin{aligned}
\sum_{p=1}^{N} z_p &= 0, \\
\sum_{p=1}^{N} z_p^2 &= N.
\end{aligned}
\end{equation}
The transmitted sequence is generated by a uniform random permutation without replacement, yielding $x_i = \Pi(z_i)$. The periodic autocorrelation function $R_x(\Delta)$ and the normalized Peak Sidelobe Level (PSL) are defined as:
\begin{equation}
R_x(\Delta) = \sum_{i=1}^{N} x_i x_{i+\Delta}
\end{equation}

For the zero-shift autocorrelation ($\Delta = 0$), the interleaver only permutes positions, yielding a deterministic main lobe representing the total energy:
\begin{equation}
\begin{aligned}
R_x(0) &= \sum_{i=1}^{N} x_i^2 = \sum_{p=1}^{N} z_p^2 = N.
\end{aligned}
\end{equation}

For any non-zero shift $\Delta \neq 0$, sampling $x_i$ and $x_{i+\Delta}$ without replacement from $\mathcal{Z}$ yields the sidelobe expectation:
\begin{equation}
\begin{aligned}
\mathbb{E}_{\Pi}[R_x(\Delta)] &= \sum_{i=1}^{N} \mathbb{E}_{\Pi}[x_i x_{i+\Delta}] \\
&= N \cdot \left( \frac{1}{N(N-1)} \sum_{p \neq q} z_p z_q \right) \\
&= \frac{1}{N-1} \left[ \left( \sum_{p=1}^{N} z_p \right)^2 - \sum_{p=1}^{N} z_p^2 \right] \\
&= -\frac{N}{N-1}.
\end{aligned}
\end{equation}

The sidelobe variance is split into the squared-expectation (Part B) and second-moment (Part A) terms:
\begin{equation}
\text{Var}[R_x(\Delta)] = \underbrace{\mathbb{E}[R_x(\Delta)^2]}_{\text{Part A}} - \underbrace{\left(\mathbb{E}[R_x(\Delta)]\right)^2}_{\text{Part B}}.
\end{equation}

Part B evaluates as:
\begin{equation}
\begin{aligned}
\lim_{N \to \infty}\left(\mathbb{E}[R_x(\Delta)]\right)^2
&= \lim_{N \to \infty}\left(-\frac{N}{N-1}\right)^2 = 1.
\end{aligned}
\end{equation}

Part A expands into a double sum of $N^2$ cross products, decomposed into three cases based on index overlap:
\begin{equation}
\mathbb{E}[R_x(\Delta)^2] = \sum_{i=1}^{N} \sum_{j=1}^{N} \mathbb{E}[x_i x_{i+\Delta} x_j x_{j+\Delta}].
\end{equation}

\textit{Case 1: Exact Overlap ($i = j$)}. Yielding $N$ terms with 2 distinct elements, the algebraic expansion gives:
\begin{equation}
\begin{aligned}
\sum_{i=1}^{N} \mathbb{E}[x_i^2 x_{i+\Delta}^2] &= N \cdot \left( \frac{\sum_{p \neq q} z_p^2 z_q^2}{N(N-1)} \right) \\
&= \frac{N^2 - \sum_{p=1}^{N} z_p^4}{N-1}.
\end{aligned}
\end{equation}

To evaluate the fourth-order term $\sum_{p=1}^{N} z_p^4$, the physical properties of the structurally whitened semantic features are invoked. It is assumed that the maximum absolute value among all feature elements is strictly bounded by a finite constant $M > 0$, independent of $N$, such that $\max_{1 \le p \le N} |z_p| \le M$.

Applying this uniform boundedness condition alongside the exact energy normalization $\sum_{p=1}^N z_p^2 = N$, the fourth-order sum is rigorously upper-bounded by factoring out the maximum squared amplitude:
\begin{equation}
\begin{aligned}
\sum_{p=1}^{N} z_p^4 &\le \max_{1 \le p \le N} (z_p^2) \sum_{p=1}^{N} z_p^2 \\
&\le M^2 \cdot N.
\end{aligned}
\end{equation}
Since $M$ is a finite constant, the sum of the fourth powers scales at most linearly with $N$. Expressed in strict asymptotic notation, the empirical sum evaluates to:
\begin{equation}
\sum_{p=1}^{N} z_p^4 = \mathcal{O}(N).
\end{equation}

Substituting this linear scaling back into the exact overlap expression, the fourth-order term acts as a lower-order quantity compared to $N^2$. The asymptotic limit yields an infinite quantity of order $\mathcal{O}(N)$:
\begin{equation}
\begin{aligned}
\lim_{N \to \infty} \sum_{i=1}^{N} \mathbb{E}[x_i^2 x_{i+\Delta}^2] &= \lim_{N \to \infty} \frac{N^2 - \mathcal{O}(N)}{N-1} \\
&= N.
\end{aligned}
\end{equation}

\textit{Case 2: Partial Overlap ($j = i \pm \Delta$)}. Yielding $2N$ terms with 3 distinct elements, the zero-mean expansion gives:
\begin{equation}
\begin{aligned}
\sum_{j = i \pm \Delta} \mathbb{E}[x_i x_{i+\Delta} x_j x_{j+\Delta}] &= 2N \cdot \left( \frac{\sum_{p \neq q \neq r} z_p z_q^2 z_r}{N(N-1)(N-2)} \right) \\
&= \frac{2N \cdot \left(2\sum_{p=1}^N z_p^4 - N^2\right)}{N(N-1)(N-2)}.
\end{aligned}
\end{equation}
Taking the limit $N \to \infty$ and substituting $\sum_{p=1}^N z_p^4 = \mathcal{O}(N)$ evaluates to a bounded constant $\mathcal{O}(1)$:
\begin{equation}
\begin{aligned}
\lim_{N \to \infty} \sum_{j = i \pm \Delta} \mathbb{E}[x_i x_{i+\Delta} x_j x_{j+\Delta}] &= \lim_{N \to \infty} \frac{-2N^3 + \mathcal{O}(N^2)}{N^3 - 3N^2 + 2N} \\
&= -2.
\end{aligned}
\end{equation}

\textit{Case 3: No Overlap ($i \neq j, j \neq i \pm \Delta$)}. Yielding $N(N-3)$ terms with 4 distinct elements, Newton-Girard identities provide the exact expression:
\begin{equation}
\begin{aligned}
\sum_{\text{No Overlap}} \mathbb{E}[x_i x_{i+\Delta} x_j x_{j+\Delta}] &= N(N-3) \cdot \left( \frac{\sum_{p \neq q \neq r \neq s} z_p z_q z_r z_s}{N(N-1)} \right. \\
&\left. \cdot \frac{1}{(N-2)(N-3)} \right) \\
&= \frac{3N^2 - 6\sum_{p=1}^N z_p^4}{(N-1)(N-2)}.
\end{aligned}
\end{equation}
Similarly, substituting the asymptotic scaling evaluates to $\mathcal{O}(1)$:
\begin{equation}
\begin{aligned}
\lim_{N \to \infty} \sum_{\text{No Overlap}} \mathbb{E}[x_i x_{i+\Delta} x_j x_{j+\Delta}] &= \lim_{N \to \infty} \frac{3N^2 - \mathcal{O}(N)}{N^2 - 3N + 2} \\
&= 3.
\end{aligned}
\end{equation}

Aggregating the three spatial cases and the squared-expectation term, Case 1 dominates as $\mathcal{O}(N)$, while all other terms are bounded by $\mathcal{O}(1)$:
\begin{equation}
\begin{aligned}
\lim_{N \to \infty} \text{Var}[R_x(\Delta)]
&= \lim_{N \to \infty} \mathbb{E}[R_x(\Delta)^2] - \left( \mathbb{E}[R_x(\Delta)] \right)^2 \\
&= \mathcal{O}(N) -2 + 3 - 1 \\
&= \mathcal{O}(N).
\end{aligned}
\end{equation}
This confirms that finite constant terms are negligible in the limit, ensuring the sidelobe variance scales linearly with $N$.

By the Combinatorial Central Limit Theorem (CCLT), the sidelobe converges asymptotically to a Gaussian distribution. Applying Extreme Value Theory (EVT), the expected maximum absolute sidelobe is bounded by:
\begin{equation}
\begin{aligned}
\max_{\Delta \neq 0} |R_x(\Delta)| &\le |\mathbb{E}[R_x(\Delta)]| + \sqrt{2 \text{Var}[R_x(\Delta)] \ln(N-1)} \\
&= \mathcal{O}(1) + \sqrt{2 \mathcal{O}(N) \ln N} \\
&= \mathcal{O}(\sqrt{N \ln N}).
\end{aligned}
\end{equation}

Finally, substituting $R_x(0) = N$ and the extreme sidelobe bound into the PSL definition, the limit as $N \to \infty$ is:
\begin{equation}
\begin{aligned}
\lim_{N \to \infty} \text{PSL} &= \lim_{N \to \infty} \frac{\max_{\Delta \neq 0} |R_x(\Delta)|}{R_x(0)} \\
&= \lim_{N \to \infty} \frac{\mathcal{O}(\sqrt{N \ln N})}{N} \\
&= 0.
\end{aligned}
\end{equation}
Thus, pseudo-random interleaving completely shatters the structural correlation of semantic features, driving the PSL to exactly zero and mathematically ensuring asymptotic orthogonality.

\subsection{Asymptotic Orthogonality and Collision Mitigation}
In a multi-user scenario, let $\mathbf{x}_k = \Pi_k(\tilde{\mathbf{z}}_k)$ and $\mathbf{x}_j = \Pi_j(\tilde{\mathbf{z}}_j)$ denote the interleaved signals for the target user $k$ and user $j$. To quantify the semantic collision intensity in the high-dimensional feature space, the Cosine Similarity metric is directly adopted. The semantic similarity $S_{j \to k}$ is defined as:
\begin{equation}\label{eq:cosine_sim_def}
S_{j \to k} = \frac{\langle \mathbf{x}_k, \mathbf{x}_j \rangle}{\|\mathbf{x}_k\|_2 \|\mathbf{x}_j\|_2}.
\end{equation}

Recall the unit-variance normalization $\frac{1}{N}\sum x_{m}^2 = 1$, which implies the $L_2$-norm of any interleaved semantic vector is deterministically $\|\mathbf{x}\|_2 = \sqrt{N}$. Thus, the cosine similarity simplifies to the normalized cross-correlation:
\begin{equation}\label{eq:cosine_sim_simplified}
S_{j \to k} = \frac{1}{N} \sum_{m=1}^{N} x_{k,m} x_{j,m}.
\end{equation}

Assuming the permutation operators $\Pi_k$ and $\Pi_j$ are generated independently, the expectation of the semantic similarity naturally vanishes due to the zero-mean property:
\begin{equation}\label{eq:cosine_mean}
\mathbb{E}[S_{j \to k}] = \frac{1}{N} \sum_{m=1}^{N} \mathbb{E}[x_{k,m}] \mathbb{E}[x_{j,m}] = 0.
\end{equation}

Consequently, the fluctuation of this similarity is determined by its variance:
\begin{equation}\label{eq:cosine_variance}
\begin{aligned}
\text{Var}(S_{j \to k}) &= \mathbb{E} [S_{j \to k}^2] = \frac{1}{N^2} \mathbb{E} \left[ \langle \mathbf{x}_k, \mathbf{x}_j \rangle^2 \right] \\
&\stackrel{(a)}{=} \frac{1}{N^2} \left( \sum_{m=1}^N \mathbb{E}[x_{k,m}^2] \mathbb{E}[x_{j,m}^2] \right. \\
&\left.\quad + \sum_{m \neq n} \mathbb{E}[x_{k,m} x_{k,n}] \mathbb{E}[x_{j,m} x_{j,n}] \right) \\
&\stackrel{(b)}{=} \frac{1}{N^2} \left[ N (1 \cdot 1) + N(N-1) \left( -\frac{1}{N-1} \right)^2 \right] \\
&\stackrel{(c)}{=} \frac{1}{N^2} \left( \frac{N^2}{N-1} \right) \\
&\stackrel{(d)}{=} \frac{1}{N-1},
\end{aligned}
\end{equation}
where step (b) invokes the unit variance constraint and the negative autocorrelation expectation $\mathbb{E}[x_m x_n] = -\frac{1}{N-1}$ derived from sampling without replacement.

To evaluate the probability of a severe semantic collision, we define a collision event as the absolute cosine similarity exceeding a small threshold $\epsilon$. Applying Chebyshev's inequality, this probability is strictly bounded:
\begin{equation}\label{eq:chebyshev}
\mathbb{P} \left( |S_{j \to k}| \ge \epsilon \right) \le \frac{\text{Var} (S_{j \to k})}{\epsilon^2} = \frac{1}{(N-1)\epsilon^2}.
\end{equation}

Taking the limit as the semantic dimension $N \to \infty$, we establish the asymptotic orthogonality:
\begin{equation}\label{eq:collision_prob}
\lim_{N \to \infty} \mathbb{P} \left( |S_{j \to k}| \ge \epsilon \right) = 0.
\end{equation}
This rigorous limit mathematically proves that, through independent structural whitening, multi-user semantic vectors are forced into mutually orthogonal subspaces in probability. From the decoder's perspective, the interfering semantic objects lose their structural correlation entirely, effectively averting semantic collapse and acting as manageable background noise.

\section{Simulation Results}
In this section, the performance of the proposed SIDMA system is evaluated through extensive simulations. The experiments are conducted on the \textbf{DIV2K dataset}, a high-quality image restoration benchmark. Specifically, the \textbf{800 training images} are utilized to train the Swin Transformer-based semantic transcoder and the neural network-based adaptive power allocation module, while the remaining \textbf{100 testing images} are used for performance evaluation.

To demonstrate the superiority of SIDMA, it is compared against the following state-of-the-art AI-native and traditional multiple access baseline schemes:
\begin{itemize}
    \item \textbf{OMDMA (Orthogonal Model Division Multiple Access)\cite{omdma}:} An AI-native multiple access scheme that achieves multi-user transmission by assigning distinct model components or sub-networks to different users, representing the latest paradigm in model-level orthogonalization.
    \item \textbf{DeepMA (Deep Multiple Access)\cite{zhang2024deepma}:} An end-to-end deep multiple access framework where different users' semantic features are encoded into mutually orthogonal semantic symbol vectors (SSVs) to mitigate multi-user interference, representing a paradigm of symbol-level orthogonalization.
    \item \textbf{SE (Shared Embedding)\cite{lee2026transformer}:} A Transformer-based multiple access scheme where multiple users jointly occupy a high-dimensional embedding space and are separated by learnable user-specific positional masks, representing the latest paradigm in embedding-space multiplexing.
    \item \textbf{Traditional MA:} This baseline represents conventional multiple access frameworks. It employs classical source coding methods, specifically \textbf{JPEG} and \textbf{JPEG2000}, coupled with channel coding that operates at the \textbf{Shannon capacity limit} for bit-level transmission.
\end{itemize}

The simulation results are presented in four parts: first, the seed selection mechanism is validated through a microscopic analysis of the interleaving mechanism; subsequently, the image reconstruction quality is compared across various SNR levels; furthermore, the system's scalability is investigated under increasing user density; and finally, systematic ablation studies are conducted to isolate the performance gains of the core modules.

\subsection{Validation of the Seed Selection Mechanism}
The effectiveness of the permutation operator $\Pi_k$ in SIDMA hinges on its ability to suppress spatial correlation. While it has been theoretically proven that random permutations achieve asymptotic whitening as $N \to \infty$, in practical systems with finite feature dimensions, different interleaving seeds lead to varying levels of residual structural correlation. To demonstrate this, the Peak Sidelobe Level (PSL) matrices of semantic features under Random Seeds and Selected Seeds are compared.

The optimization goal for seed selection is to find an optimal permutation $\Pi^*$ that minimizes the maximum cross-correlation sidelobe:
\begin{equation}
\Pi^* = \arg \min_{\Pi \in \mathcal{P}} \left\{ \max_{\Delta \neq 0} |R_{\Pi}(\Delta)| \right\},
\end{equation}
where $\mathcal{P}$ is the set of candidate permutations and $R_{\Pi}(\Delta)$ is the autocorrelation function of the interleaved features.

\begin{figure*}[t]
\centering
\subfloat[Random Seed]{\includegraphics[width=0.48\linewidth]{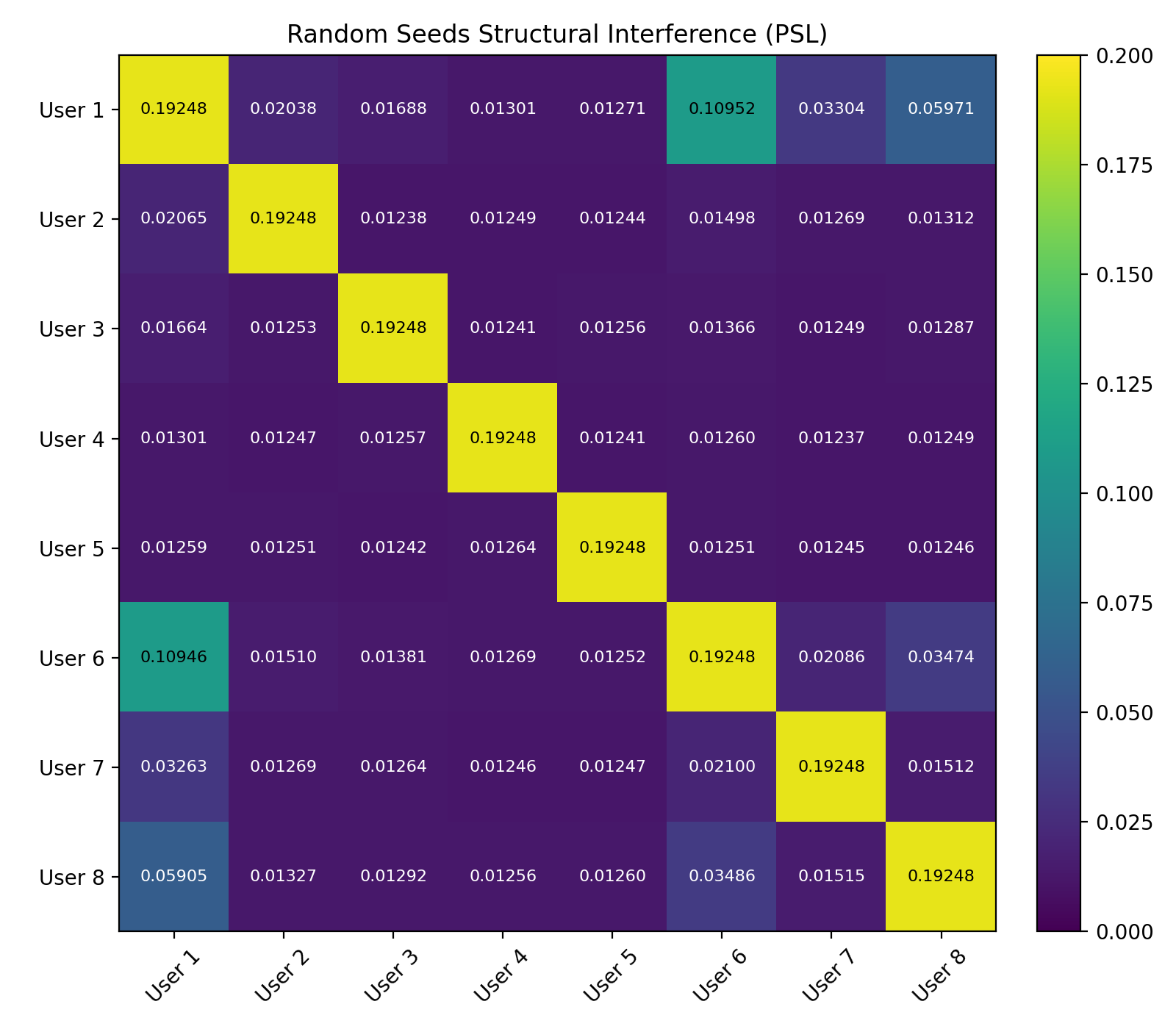}\label{fig:heatmap_random}}
\hfill
\subfloat[Selected Seed]{\includegraphics[width=0.48\linewidth]{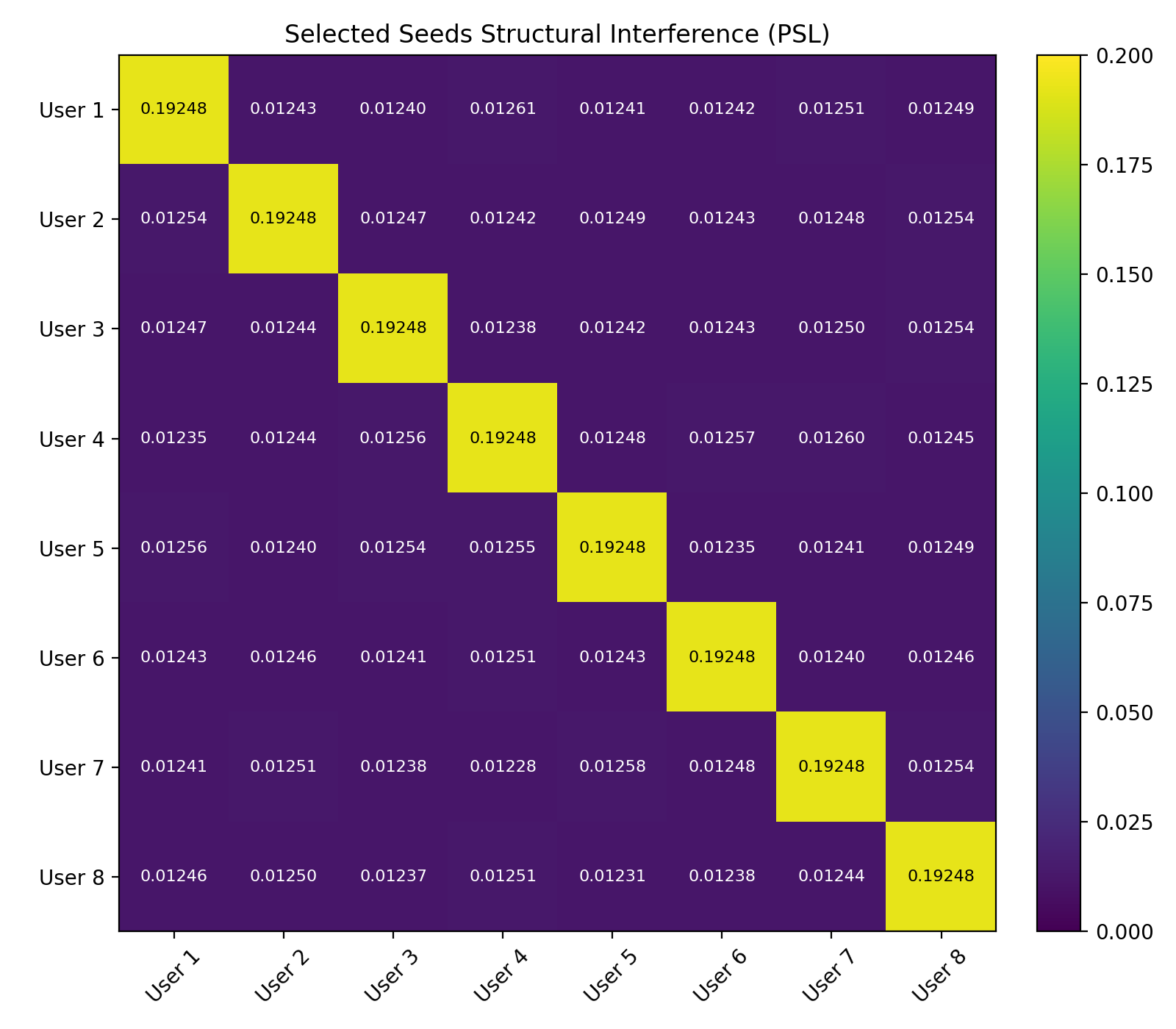}\label{fig:heatmap_selected}}
\caption{Comparison of PSL matrices (autocorrelation heatmaps) for semantic features. (a) Under a random interleaving seed, visible sidelobes indicate residual spatial correlation. (b) Under a selected seed, the sidelobes are significantly suppressed, achieving a near-ideal ``white'' distribution.}
\label{fig:psl_comparison}
\end{figure*}

Fig. \ref{fig:psl_comparison} illustrates the autocorrelation heatmaps of the de-interleaved domain for both cases. In Fig. \ref{fig:heatmap_random}, when an arbitrary random seed is used, the heatmap exhibits several distinct ``hot spots'' in the non-central regions. These peaks represent residual structural similarities where the permutation failed to completely decorrelate the semantic features, potentially leading to semantic collisions during multi-user multiplexing.

In contrast, as shown in Fig. \ref{fig:heatmap_selected}, the use of a selected seed—\textbf{generated via Algorithm 1}—yields a significantly more uniform heatmap. The off-center energy is effectively scattered, and the PSL is significantly reduced.

This ``cleaned'' background confirms the validity of the seed selection mechanism described in Section III-D, \textbf{thereby thoroughly demonstrating the practical effectiveness of Algorithm 1}. By selecting seeds that minimize the $PSL$, the SIDMA system can more effectively transform structured multi-user interference into non-coherent noise, thereby providing a more robust foundation for the subsequent joint decoding and image reconstruction process.

\subsection{Performance Evaluation Across Different SNR Levels}
\begin{figure*}[!t]
\centering
\subfloat[PSNR comparison]{\includegraphics[width=0.48\textwidth]{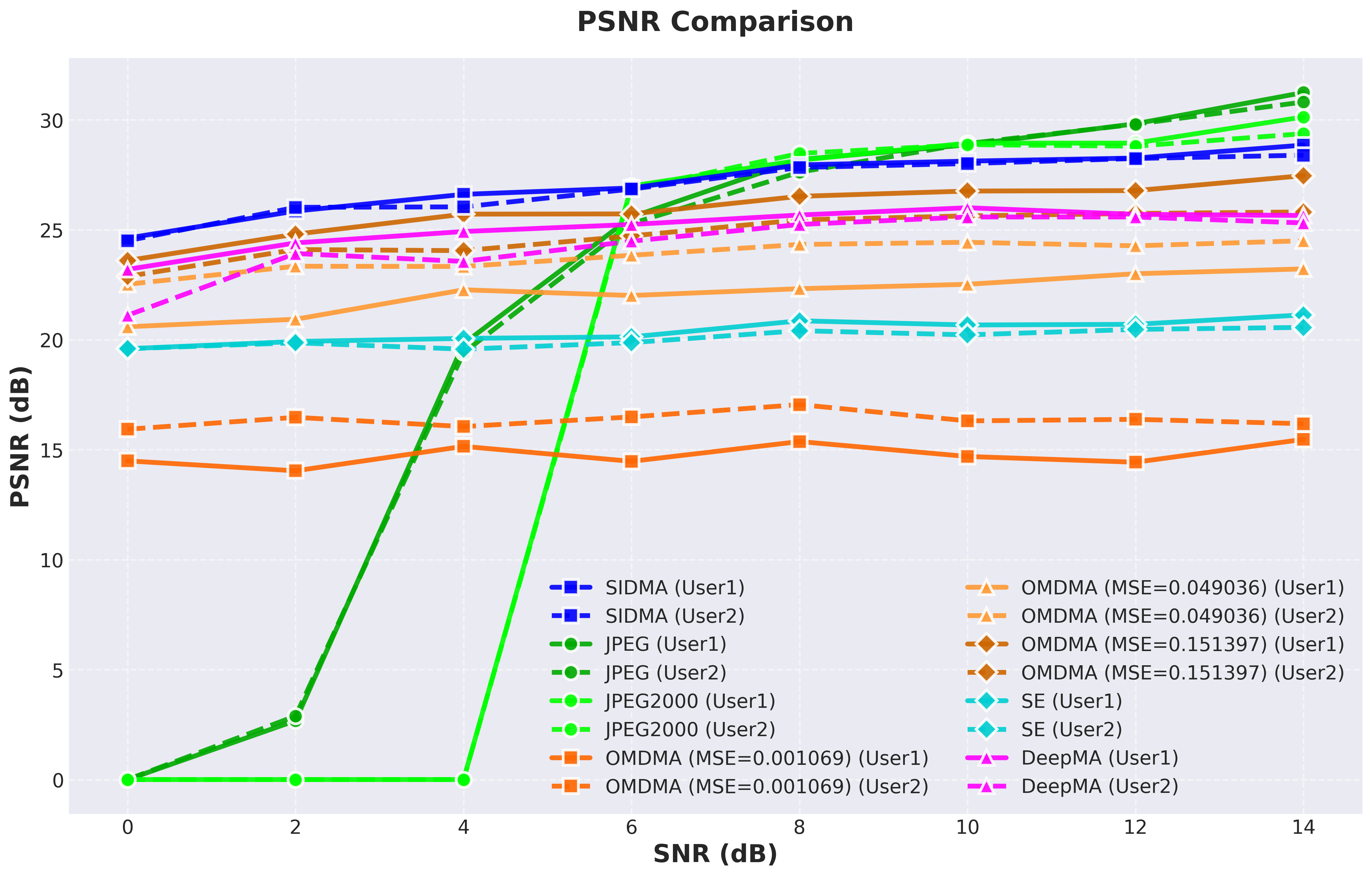}\label{fig:psnr_comparison}}\hfill
\subfloat[SSIM comparison]{\includegraphics[width=0.48\textwidth]{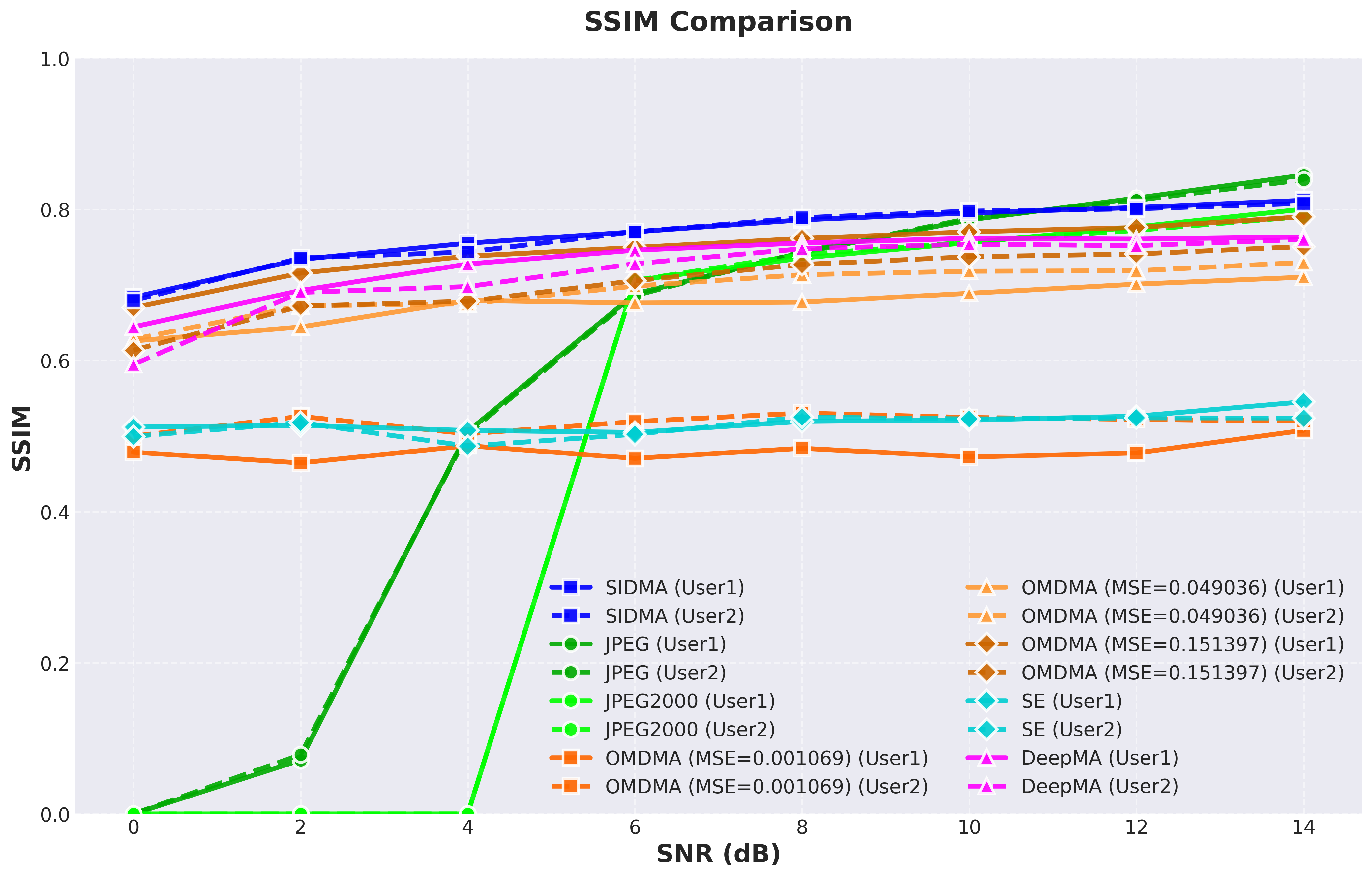}\label{fig:ssim_comparison}}
\caption{Performance comparison of SIDMA, DeepMA, SE, OMDMA, and traditional MA schemes in terms of average PSNR and SSIM versus SNR on the DIV2K test set ($K=2$ users). Note that the Mean Squared Error (MSE) values appended to the OMDMA labels indicate the degree of differentiation between the models assigned to different users.}\label{fig:sidma_vs_omdma}
\end{figure*}
To quantitatively evaluate the quality of the reconstructed images, the Peak Signal-to-Noise Ratio (PSNR) and the Structural Similarity Index (SSIM) are adopted as the core performance metrics. The PSNR is defined as follows:
\begin{equation}
\text{PSNR} = 10 \log_{10} \left( \frac{\text{MAX}_I^2}{\text{MSE}} \right),
\end{equation}
where $\text{MAX}_I$ denotes the maximum possible pixel value of the image, and $\text{MSE}$ represents the mean squared error between the original and reconstructed images. The SSIM is utilized to measure the similarity between two images in terms of luminance, contrast, and structure, which is defined as:
\begin{equation}
\text{SSIM}(x, y) = \frac{(2\mu_x \mu_y + c_1)(2\sigma_{xy} + c_2)}{(\mu_x^2 + \mu_y^2 + c_1)(\sigma_x^2 + \sigma_y^2 + c_2)},
\end{equation}
where $\mu$ and $\sigma$ represent the mean and standard deviation of the image pixels, respectively, $\sigma_{xy}$ is the covariance, and $c_1$ and $c_2$ are constants to stabilize the division.

Fig. \ref{fig:sidma_vs_omdma} illustrates the average PSNR and SSIM as functions of the SNR. The image reconstruction performance of the proposed SIDMA architecture is evaluated against advanced semantic multiple access baselines (DeepMA, SE, OMDMA) and traditional multiple access schemes (JPEG, JPEG2000) over an Additive White Gaussian Noise (AWGN) channel across varying SNR levels. Notably, the Mean Squared Error (MSE) values appended to the OMDMA labels indicate the degree of differentiation between the models assigned to different users. This differentiation metric is calculated by cross-evaluating a subset of test images---specifically, passing the latent features from one model's encoder into the other model's decoder---and averaging the MSE between these cross-reconstructed outputs and the original images. This exposes a critical limitation of the OMDMA paradigm: when the models lack sufficient differentiation, the algorithm struggles to isolate user features, leading to severe semantic collisions and degraded multiple access transmission performance.

Across the entire SNR range (0 dB to 14 dB), SIDMA consistently achieves the highest PSNR and SSIM scores. Among the baseline schemes, DeepMA delivers the best performance, followed by SE and OMDMA, while SIDMA maintains a performance margin over all of them. Particularly in the low SNR regime (e.g., 0 dB), SIDMA achieves a significant metric lead, obtaining a PSNR of approximately 24.5 dB and an SSIM of 0.65, outperforming DeepMA (22 dB), SE (20 dB), and OMDMA (23.5 dB). This demonstrates that although DeepMA (symbol-level), SE (embedding-level), and OMDMA (model-level) attempt to establish orthogonal representation spaces through distinct isolation mechanisms, their strict structural isolation is highly vulnerable to severe channel impairments. In contrast, SIDMA breaks the spatial structural correlation of features, transforming structured multi-user interference into a manageable unstructured white noise floor, thereby exhibiting greater robustness.

Compared with the semantic multiple access schemes, the curves for traditional MA schemes (JPEG and JPEG2000) clearly illustrate a catastrophic ``cliff effect.'' In the low SNR range (0--5 dB), the performance of traditional schemes drops sharply, leading to a total failure in image reconstruction. Conversely, SIDMA exhibits graceful degradation; by flattening dense semantic collisions into an unstructured background, the Swin Transformer-based decoder can effectively perform denoising and recover core semantic features even under extreme noise conditions. This comprehensive superiority serves as direct evidence for the effectiveness of the core modules in SIDMA. Specifically, by utilizing the optimal permutation seeds generated via Algorithm 1, the residual spatial correlation is minimized to achieve an excellent whitening effect. Concurrently, the integration of the neural network-based ImpPA module ensures that critical semantic elements receive differentiated protection, thereby maximizing the effective SINR for essential features.

\subsection{Impact of the Number of Users on Performance}
\begin{figure*}[!t]
\centering
\subfloat[PSNR at SNR = 0 dB\label{fig:psnr_0db}]{%
\includegraphics[width=0.48\linewidth]{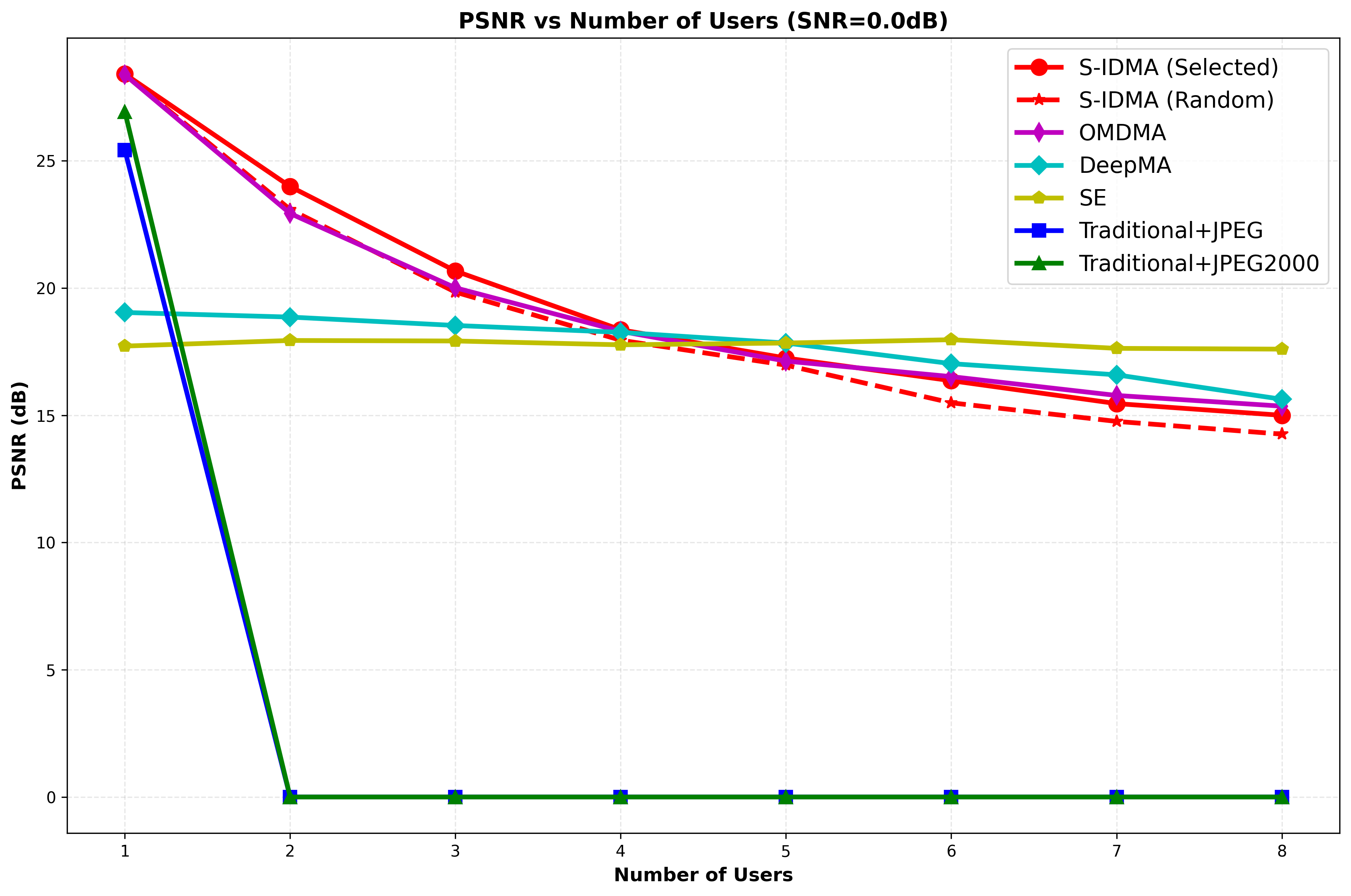}}
\hfill
\subfloat[SSIM at SNR = 0 dB\label{fig:ssim_0db}]{%
\includegraphics[width=0.48\linewidth]{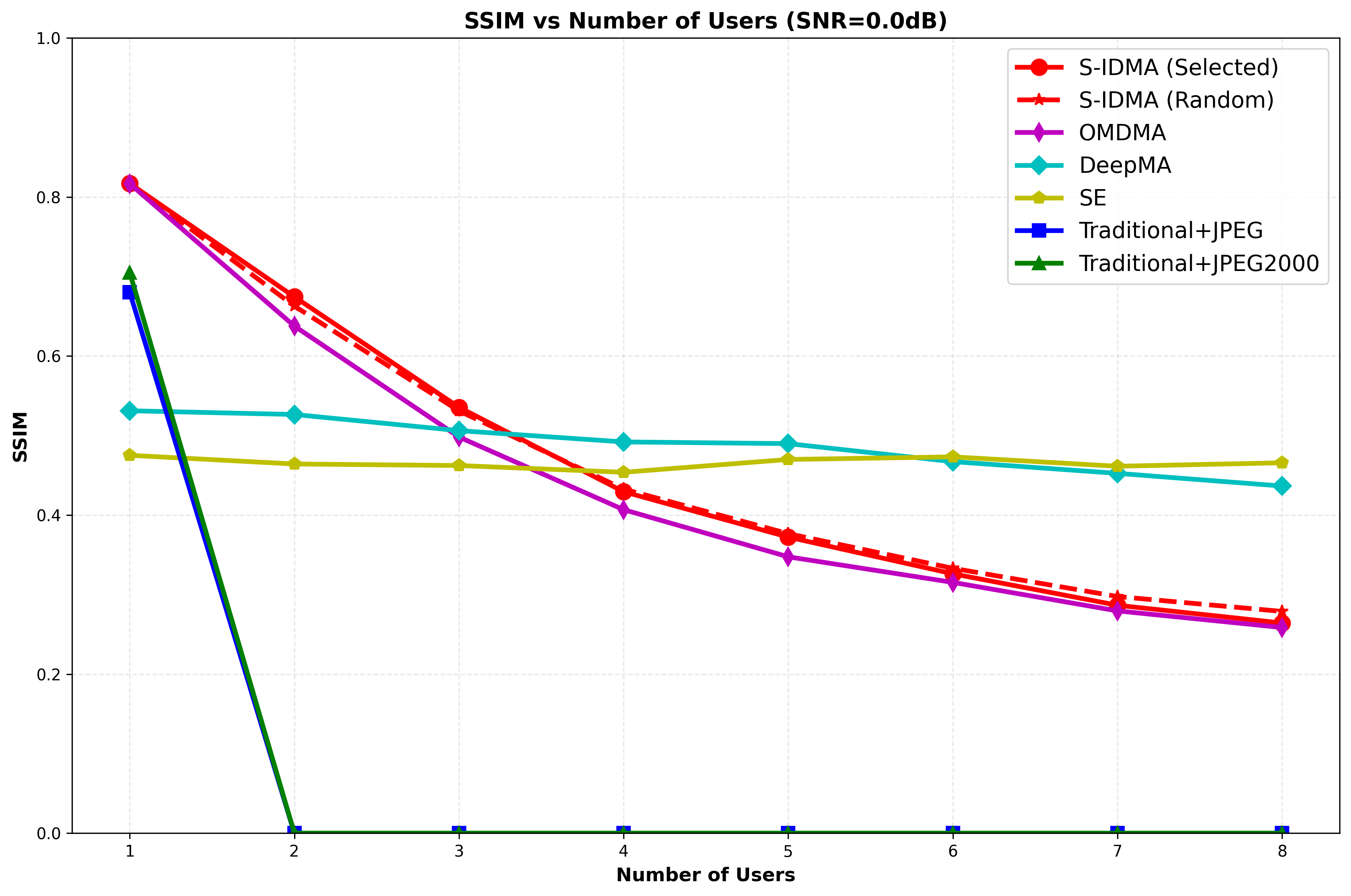}}\\[1ex]
\subfloat[PSNR at SNR = 5 dB\label{fig:psnr_5db}]{%
\includegraphics[width=0.48\linewidth]{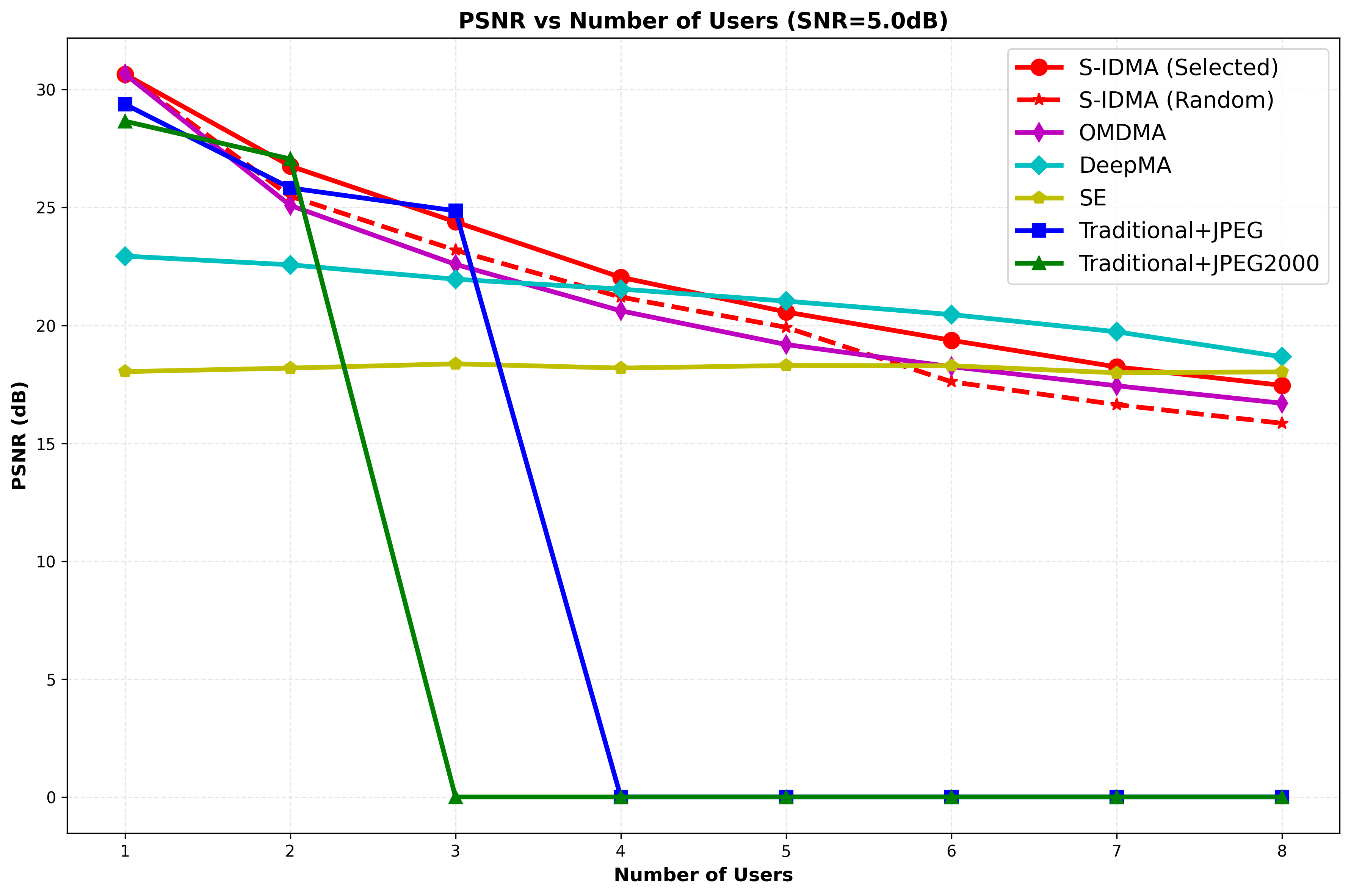}}
\hfill
\subfloat[SSIM at SNR = 5 dB\label{fig:ssim_5db}]{%
\includegraphics[width=0.48\linewidth]{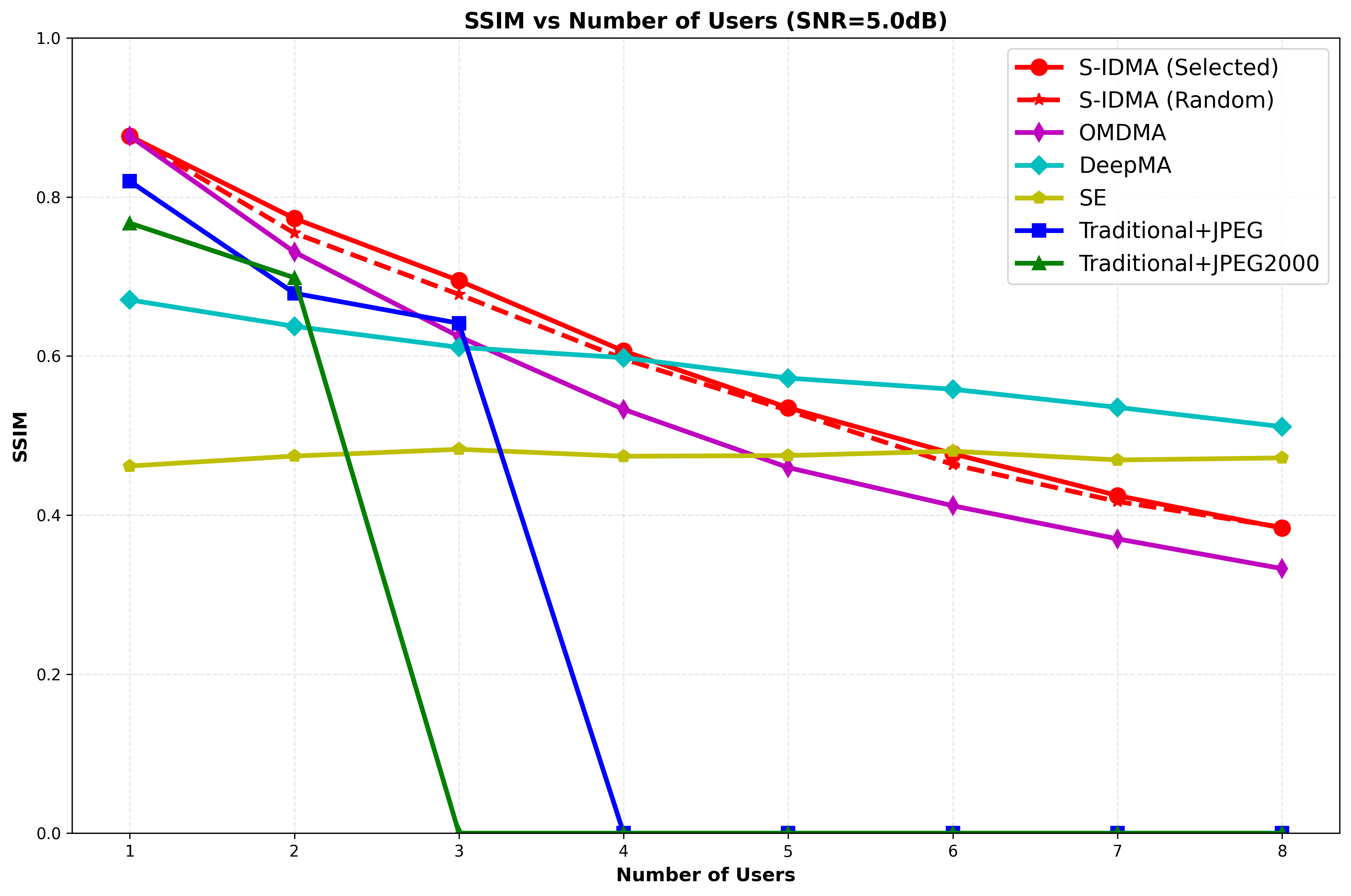}}\\[1ex]
\subfloat[PSNR at SNR = 10 dB\label{fig:psnr_10db}]{%
\includegraphics[width=0.48\linewidth]{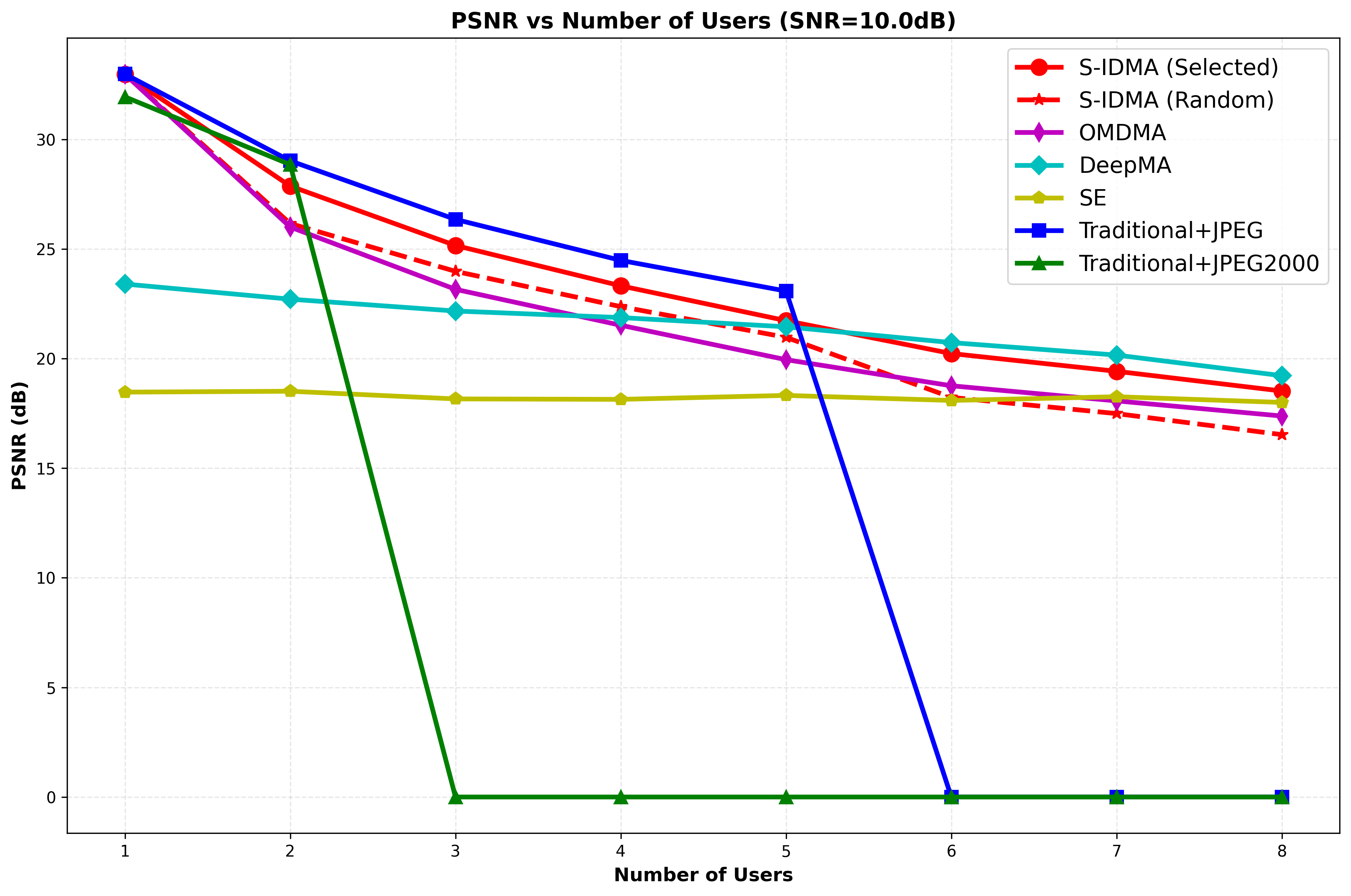}}
\hfill
\subfloat[SSIM at SNR = 10 dB\label{fig:ssim_10db}]{%
\includegraphics[width=0.48\linewidth]{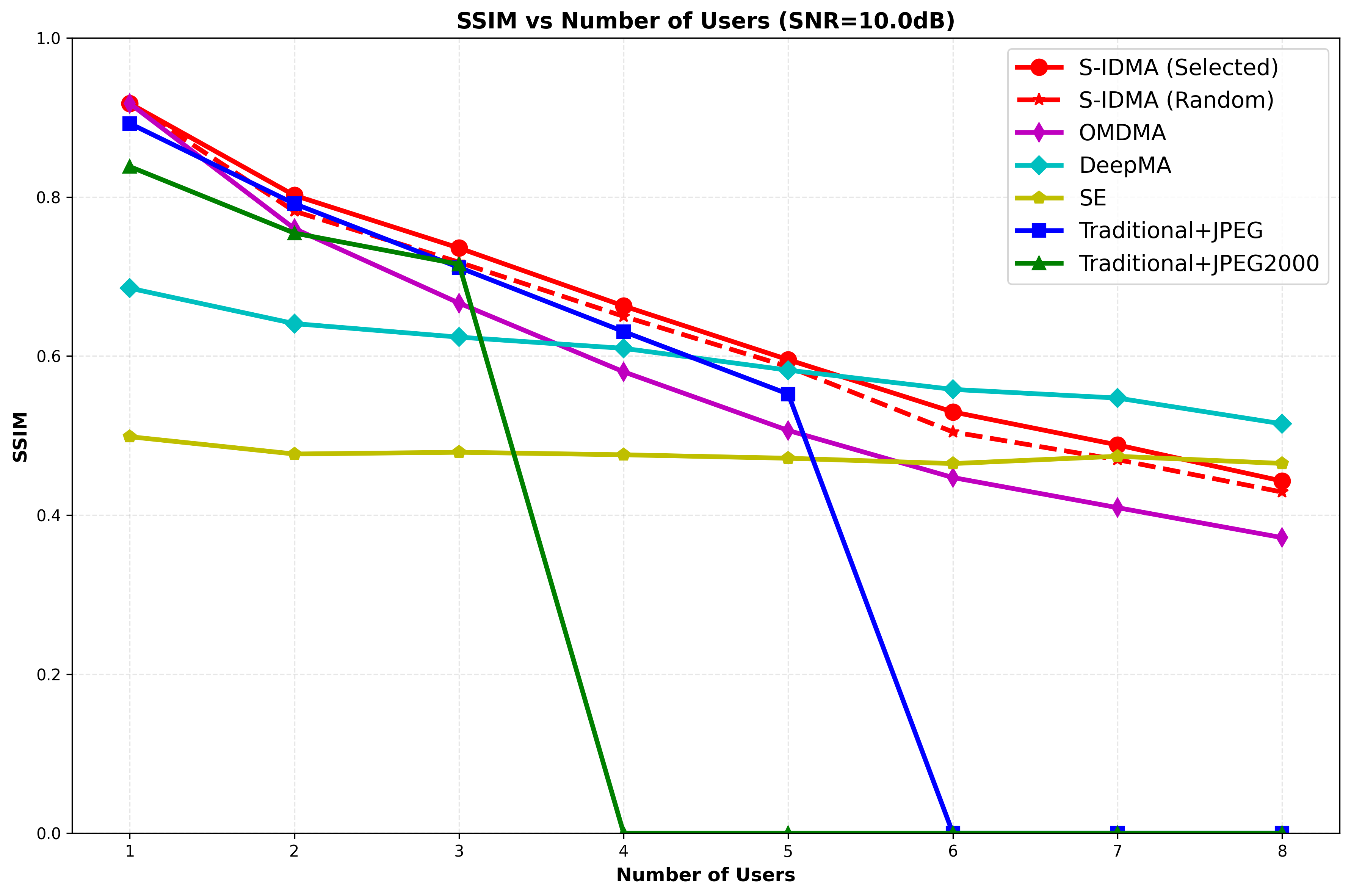}}
\caption{Performance comparison of multiple access schemes under varying user densities and SNR levels. The results reveal a clear operational trade-off: SIDMA achieves significantly higher reconstruction fidelity at low to moderate user densities, whereas DeepMA and SE exhibit flatter decay curves under extreme multi-user congestion, while OMDMA and traditional MA suffer from rapid degradation.}
\label{fig:multi_user_visual}
\end{figure*}

Fig. \ref{fig:multi_user_visual} evaluates the system scalability by plotting the average PSNR and SSIM against an increasing number of multiplexed users ($K=1$ to $8$) over an AWGN channel. A prominent observation is the distinct operational regimes where the proposed SIDMA and existing semantic multiple access algorithms excel. When the number of users is relatively small, SIDMA exhibits a performance advantage over all evaluated baselines. By breaking spatial structural correlation and transforming interference into an unstructured noise floor, SIDMA achieves exceptional reconstruction fidelity, outperforming the model-level, symbol-level, and embedding-level orthogonalization strategies of OMDMA, DeepMA, and SE, respectively. In this regime, SIDMA fully exploits the available channel resources to deliver high-quality semantic transmission.

However, as the user density increases significantly, the performance dynamics shift. The reconstruction quality of SIDMA experiences a more noticeable decline due to the overwhelming accumulation of non-orthogonal interference energy, eventually falling behind DeepMA and SE in highly congested scenarios. In contrast, OMDMA not only fails to achieve the high fidelity of SIDMA at low user counts but also suffers from a rapid performance degradation as user density increases, indicating that its model-level orthogonalization is highly vulnerable to severe semantic collisions. While DeepMA and SE suffer from a severely low absolute performance ceiling at low user counts (indicating an inability to achieve high-fidelity transmission), their strict symbol-level and embedding-level isolation mechanisms demonstrate flatter decay curves. This makes them more robust when a large number of users are densely multiplexed. Furthermore, traditional MA baselines (JPEG and JPEG2000) consistently suffer from a catastrophic ``cliff effect,'' failing completely once the rigid channel capacity is exceeded.

Ultimately, these simulation results objectively delineate the optimal application scenarios for different paradigms. SIDMA provides an unparalleled, high-fidelity multiple-access solution for small to moderate user densities, effectively bridging the gap between semantic extraction and physical-layer multiplexing. Meanwhile, algorithms like DeepMA and SE offer a trade-off, sacrificing peak reconstruction quality for better extensibility under extreme multi-user congestion, whereas rigid partitioning schemes like OMDMA and traditional MA fail to sustain performance in dense networks.

\subsection{Ablation Analysis}
\begin{figure}[!t]
\centering
\subfloat[PSNR performance with ImpPA\label{fig:imppa_psnr}]{%
\includegraphics[width=0.90\linewidth]{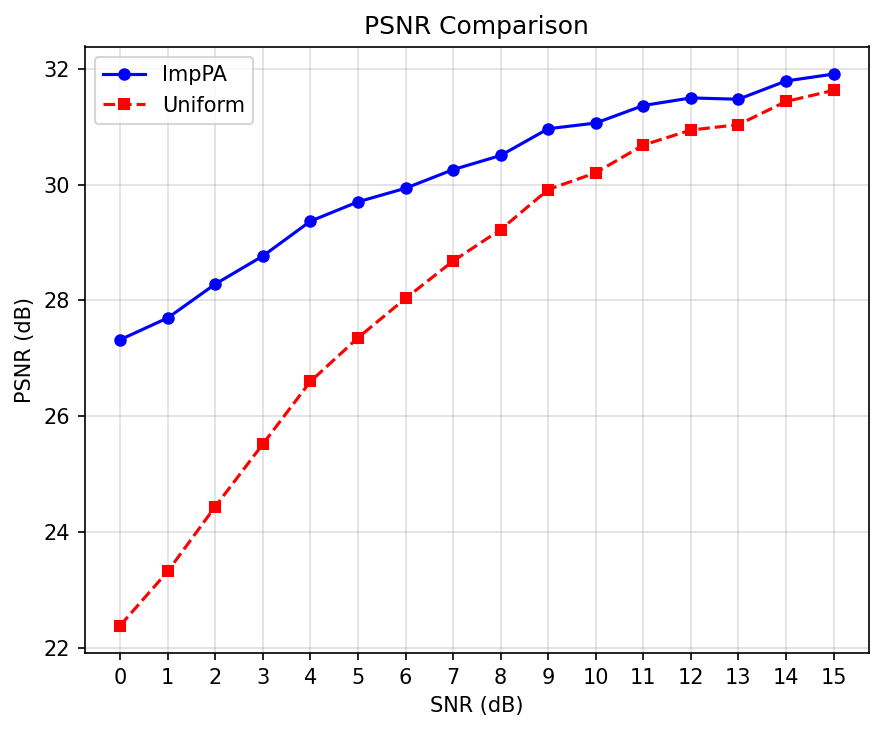}}
\hfill
\subfloat[SSIM performance with ImpPA\label{fig:imppa_ssim}]{%
\includegraphics[width=0.90\linewidth]{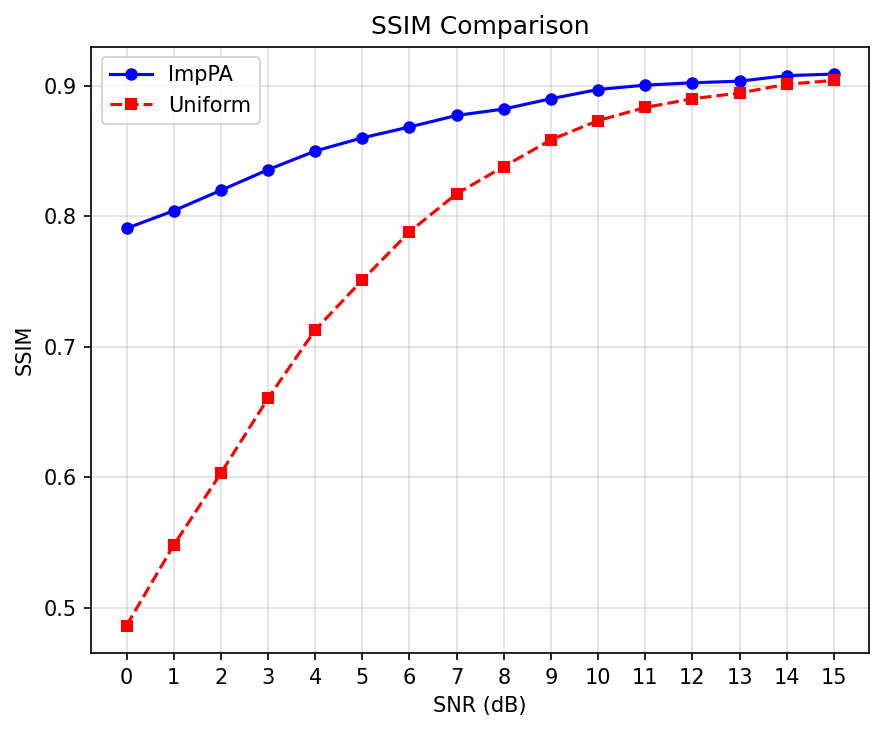}}
\caption{Ablation study comparing ImpPA-based adaptive power allocation with uniform power allocation across various SNR levels. (a) PSNR comparison demonstrating the reconstruction quality gain. (b) SSIM comparison illustrating the structural similarity enhancement provided by the ImpPA module.}
\label{fig:imppa_ablation}
\end{figure}

This subsection presents an ablation study evaluating the Importance-aware Power Allocation (ImpPA) module by comparing the complete SIDMA architecture against a uniform power allocation baseline. As illustrated in Fig. \ref{fig:imppa_ablation}, the ImpPA module demonstrates significant superiority across the 0–15 dB SNR range. Notably, at 0 dB, ImpPA achieves a PSNR of 27.3 dB (a 5 dB gain) and an SSIM of 0.79 (compared to 0.49 for the uniform baseline), proving its effectiveness in prioritizing critical semantic features. This gain indicates that by concentrating transmission energy on essential semantic blocks, the system effectively mitigates interference and prevents ``semantic collapse.'' As SNR increases, performance curves converge, suggesting a reduced reliance on aggressive scheduling under favorable channel conditions. These results validate the ImpPA module as a robust solution for securing core semantic reconstruction in resource-constrained multi-user environments.

\begin{figure}[t]
\centering
\subfloat[PSNR performance.\label{fig:psnr_ablation}]{%
\includegraphics[width=0.90\linewidth]{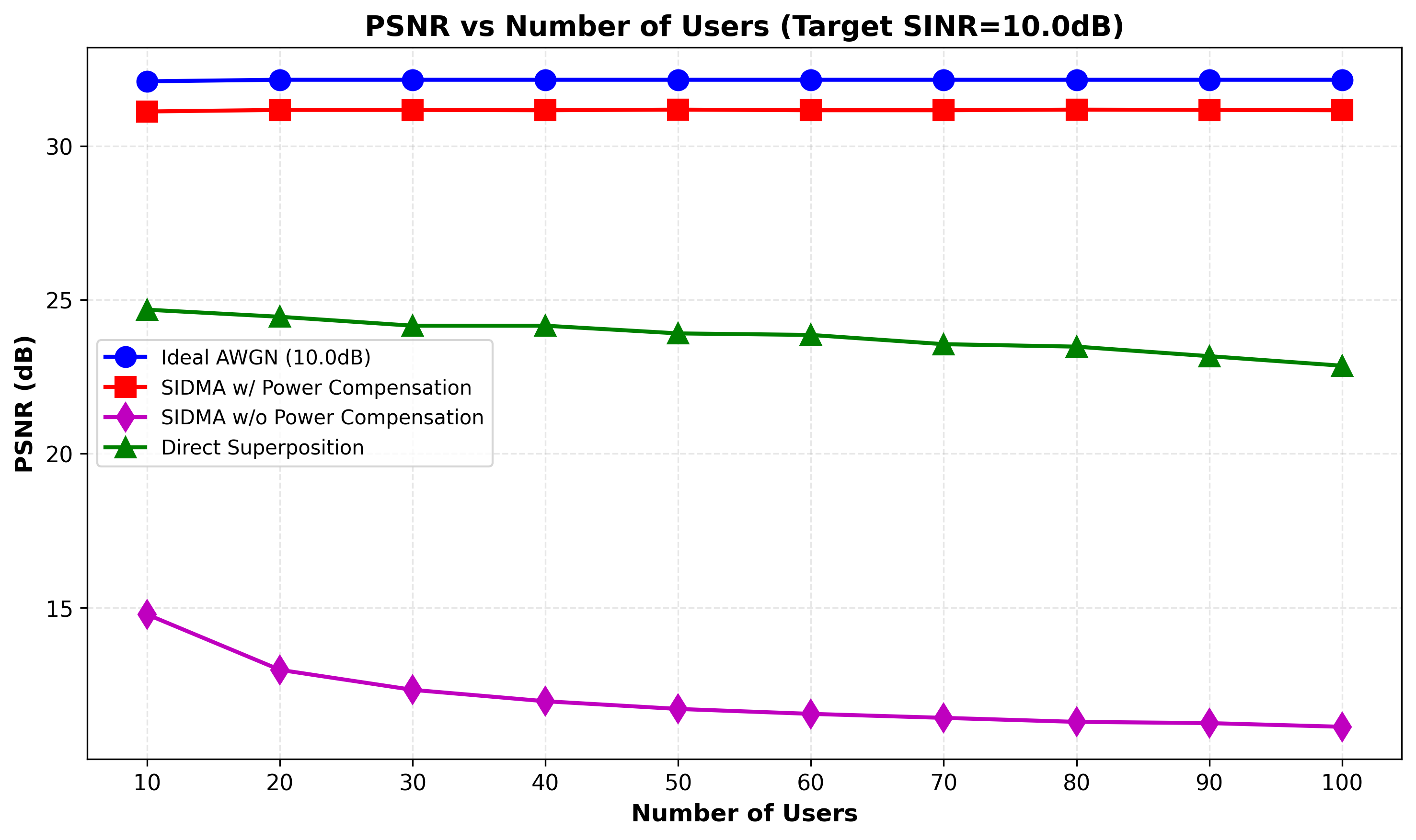}}\\[1ex]
\subfloat[SSIM performance.\label{fig:ssim_ablation}]{%
\includegraphics[width=0.90\linewidth]{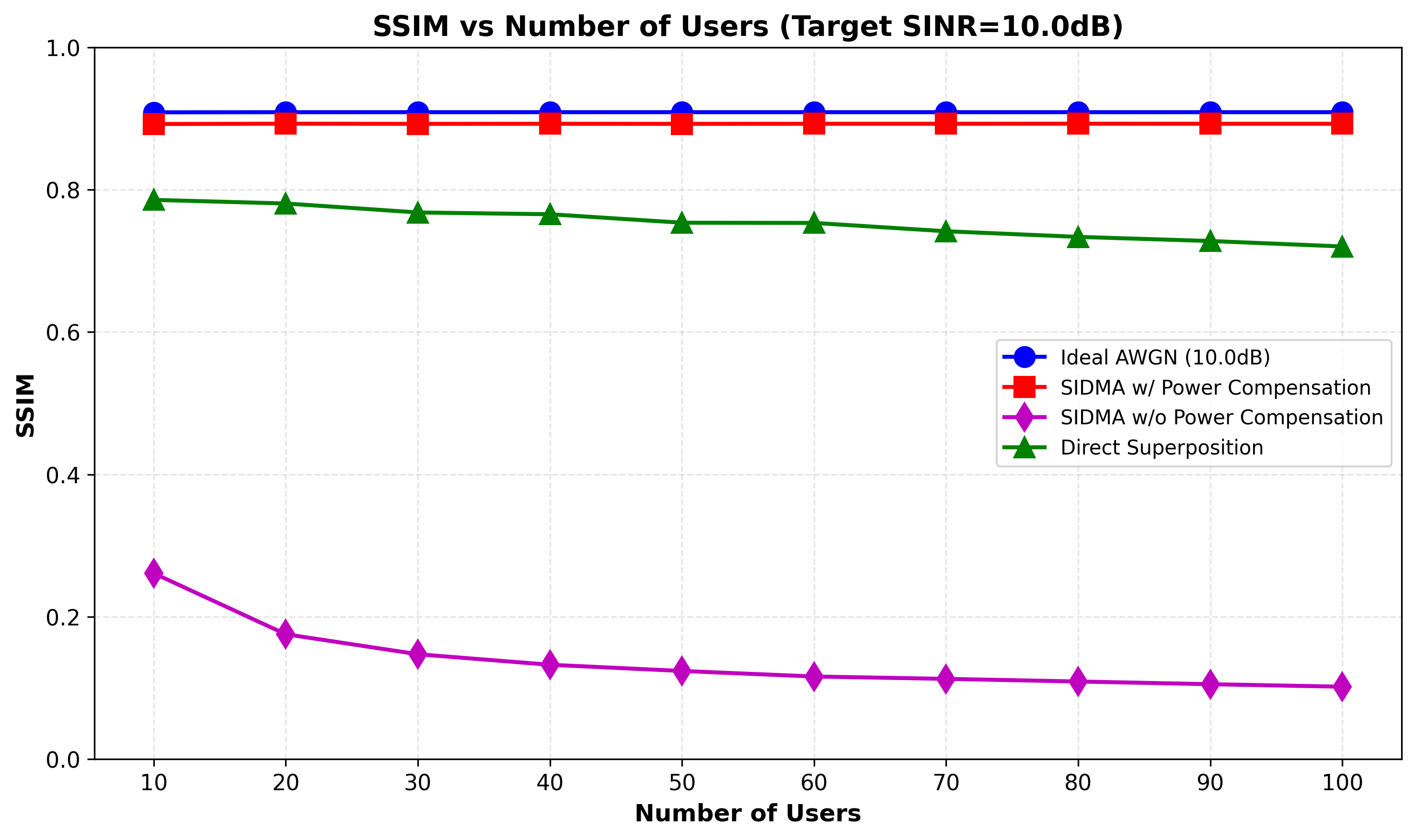}}
\caption{Ablation study evaluating the synergy between structural whitening and power compensation in SIDMA at a target SINR of 10.0 dB across varying user densities ($K \in [10, 100]$). The results highlight the crucial role of semantic interleaving, which successfully transforms intractable multi-user collisions into a mathematically manageable white noise distribution. Building upon this structural whitening, power compensation effectively suppresses the uniform background noise, thereby unleashing the full potential of SIDMA to achieve near-ideal reconstruction fidelity.}
\label{fig:ablation_study}
\end{figure}

To isolate and validate the contributions of the core modules, an ablation study is conducted by scaling the multiplexed users from 10 to 100 at a target SINR of 10.0 dB. As shown in Fig. \ref{fig:ablation_study}, the complete SIDMA architecture (with power compensation) exhibits exceptional scalability, maintaining high-fidelity reconstruction that nearly overlaps with the interference-free Ideal AWGN bound even under severe multi-user congestion.

An in-depth comparison between ``SIDMA without power compensation'' and ``Direct Superposition'' reveals the core physical objective of this design. While semantic interleaving successfully achieves ``structural whitening'' by breaking up dense multi-user collisions, this dispersed energy objectively persists as an elevated white noise floor. If left unmanaged, this uniform noise completely drowns out the target signal, yielding worse performance than Direct Superposition (which still retains some decodable structural correlations). Therefore, the fundamental purpose of the power compensation is precisely to counteract and eliminate this interleaving-generated white noise. Only when this noise floor is successfully suppressed can the ``structural whitening'' advantage of interleaving be fully unleashed, fundamentally solving the multi-user collision problem and enabling near-ideal fidelity in ultra-dense networks.

\section{Conclusion}
This paper proposes a Semantic Interleave Division Multiple Access (SIDMA) system architecture. It aims to address semantic collisions and collapse in multi-user concurrent transmissions. By integrating structural whitening in the feature domain—achieved through pseudo-random interleaving—with an Importance-aware Power Allocation (ImpPA) module, SIDMA successfully transforms destructive multi-user interference into manageable noise-like signals. Simultaneously, it provides robust protection for critical semantic features. Theoretical derivations based on extreme value theory and extensive simulation results demonstrate that SIDMA establishes asymptotic orthogonality in high-dimensional feature spaces. Compared with advanced semantic multiple access schemes such as OMDMA, DeepMA, and SE, SIDMA exhibits superior reconstruction fidelity in multi-user concurrent transmissions and maintains enhanced scalability and robustness in ultra-dense networks.



\bibliographystyle{ieeetr}
\bibliography{references}


 





\end{document}